\documentclass[review]{elsarticle}

\usepackage{lineno,hyperref}
\usepackage{tikz-cd}
\usepackage{dsfont}
\usepackage{amsfonts}
\usepackage{amsmath}
\usepackage{slashed}
\usepackage{graphics}
\usepackage{amsmath}
\usepackage{amssymb}
\usepackage{tikz-cd}
\usepackage[compat=1.1.0]{tikz-feynman}
\usepackage{setspace}
\usepackage{float}
\usepackage{cleveref}
\usepackage{units}
\modulolinenumbers[5]

\journal{Symmetry (MDPI)}
\begin{document}

\begin{frontmatter}

\title{\textbf{Phenomenological Effects of CPT and Lorentz Invariance Violation in Particle and Astroparticle~Physics}}

\author[Unimi]{V. Antonelli}

\author[Unimi]{L. Miramonti}

\author[Unimi]{M.D.C. Torri\corref{mycorrespondingauthor}}
\cortext[mycorrespondingauthor]{Corresponding author}
\ead{marco.torri@unimi.it, marco.torri@mi.infn.it}

\address[Unimi]{Dipartimento di Fisica, Universit\'a degli Studi di Milano and INFN Milano\\via Celoria 16, I - 20133 Milano, Italy}

\begin{abstract}
It is well known that a fundamental theorem of Quantum Field Theory (QFT) set in flat spacetime ensures the $CPT$ invariance of the theory. This symmetry is strictly connected to the Lorentz covariance, and consequently to the fundamental structure of spacetime. Therefore it may be interesting to investigate the possibility of departure from this fundamental symmetry, since it can furnish a window to observe possible effects of a more fundamental quantum gravity theory in a ``lower energy limit''. Moreover, in the past, the inquiry of symmetry violations provided a starting point for {new physics discoveries}. A useful physical framework for this kind of search is provided by astroparticle physics, thanks to the high energy involved and to the long path travelled by particles accelerated by an astrophysical object and then revealed on Earth. Astrophysical messengers are therefore very important probes for investigating this sector, involving high energy photons, charged particles,
and neutrinos of cosmic origin. In addition, one can also study artificial neutrino beams, investigated at
accelerator experiments. Here we discuss the state of art for all these topics and some interesting new proposals, both from a theoretical and phenomenological point of view.\\
\end{abstract}

\begin{keyword}
\emph{Fundamental Spacetime Symmetries, CPT violation,Lorentz Invariance Violation, Quantum Gravity, Particle Physics, Astroparticle Physics, Cosmic Rays}
\end{keyword}

\end{frontmatter}

\setcounter{section}{0} 

\section{Introduction}
Symmetries are fundamental in every physical theory formulation, since they express the visible variations of detected results after changes in the experimental framework. For this reason, symmetries are useful and desirable for the consistency of a theory. Among all  possible physical symmetries, a particular and important role is played by space-time ones, i.e., the Lorentz/Poincar\'e invariance, the time reversal $T$, and the parity transform $P$, together with the charge conjugation $C$. Indeed spacetime represents the basic scenario where all physical processes take place. Nowadays Lorentz Invariance (LI) stands at the basis of our physical knowledge and underlies the formulation of the Quantum Field Theory (QFT) description of the Standard Model (SM), the gauge theory describing in a very successful way,  elementary particle interactions. Moreover invariance under $CPT$ symmetry, the combined transformation given by the product of $C$, $P$, and $T$ in any order, is guaranteed in flat spacetime by a fundamental result: The so-called $CPT$ {theorem} that was implicit in '51 Schwinger's work \cite{Schwinger} and was more formally proved by L\'uders \cite{Lueders} and by Pauli \cite{Pauli} a few years later. For every symmetry, it is interesting to test at which level of accuracy we can verify it and which are the conditions required to prove formally its validity. Moreover, symmetries can even guide to the discovery of new physics, through the study of the consequences of their potential violation, or, at least, modification. In the specific case of the Lorentz and CPT invariance the search for modifications of these fundamental principles is motivated by the idea that small violations of space-time symmetries could provide an experimental window on the Planck scale effects caused by a more fundamental quantum gravity~theory.

In this work we will give an introductory demonstration of the $CPT$ {theorem}, showing that this result is fundamental since it is strictly related to LI and therefore to the space-time intrinsic structure. We will start introducing the algebraic demonstration, with a review of the original Jost proof \cite{Jost}, based on axiomatic field theory formulation. Then we will consider the proof based on the Lagrangian field theory, following by Bell \cite{Bell}, L\'uders \cite{Lueders}, Pauli \cite{Pauli}, and even Schwinger \cite{Schwinger,Schwinger1}. After the introduction of the $CPT$ {theorem}, we will discuss the connection between Lorentz and $CPT$ invariance in QFT, giving a demonstration of the {Greenberg theorem}, which states that $CPT$ violation implies Lorentz Invariance Violation (LIV)~\cite{Greenberg2}. Then we discuss the implications of $CPT$ symmetry and  expose the motivations for conducing the search for possible violations of this fundamental symmetry. In the following section, we explain how it may be possible to generalize the $CPT$ {theorem} to curved spacetime and the implications of its validity or violation in the gravity sector. Finally we discuss the phenomenology that can be induced by $CPT$ violation and by LIV in the astrophysical sector, in particular in gamma-ray bursts, Cosmic Rays (CRs), and the neutrino sector.

\section{$CPT$ Theorem}
In this section we will give a review of the $CPT$ {theorem} by first introducing the axiomatic demonstration and then the Lagrangian or operator-based formulation, as discussed in \cite{Streater,Bogolyubov,Haag}. The~$CPT$ theorem states that a Lorentz invariant QFT that preserves unitarity and locality and is defined in a flat spacetime is invariant under the action of the operator $\theta=CPT$, where the product order of $C$, $P$, and $T$ does not matter. Under the action of such an operator, a generic physical state transforms as:
\begin{equation}
\label{1}
|\psi\rangle_{CPT}=\theta|\psi\rangle=|\theta\psi\rangle=|\overline{\psi}\rangle
\end{equation}
so the theory must describe particles and antiparticles and:
\begin{equation}
\label{2}
\langle\overline{\psi}|\overline{\psi}\rangle=\langle\psi|\theta^{\dagger}\theta|\psi\rangle.
\end{equation}

Since two combined charge conjugations, parity, and time reversal transforms leave  the final state unaffected, it follows that the unitary property:
\begin{equation}
\label{3}
\theta^{\dagger}\theta=\mathbb{I}\;\Rightarrow\;\theta^{-1}=\theta^{\dagger}.
\end{equation}

Now it is important to underline that $T$ is antilinear, while $C$ and $P$ are both linear transformations, so the total transformation $\theta=CPT$ is antilinear, therefore $\theta$ results antiunitary. Hence for a generic hermitian operator $A$, the $CPT$ correlated one is given by:
\begin{equation}
\label{4}
A_{CPT}=\theta A\theta^{\dagger}=\theta^{\dagger}A\theta
\end{equation}
therefore:
\begin{equation}
\label{5}
\langle\overline{\psi}|A|\overline{\chi}\rangle=\langle\psi|\theta^{\dagger}A\theta|\chi\rangle^{*}=\langle\psi|A_{CPT}|\chi\rangle^{*}=\langle\chi|A_{CPT}^{\dagger}|\psi\rangle
\end{equation}
thanks to the antiunitarity of $\theta$.

\subsection{Wightman Axioms}
The axiomatic argumentation is important since it can give an inside view on the importance of  $CPT$ symmetry caused by its strict correlation with Lorentz invariance and can underline that  complete $CPT$ symmetry presents a more fundamental nature than either $C$, $P$, or $T$ alone. Before introducing the Jost argumentation \cite{Jost}, it is first necessary to review the Wightman axioms  background required to define a QFT on a flat spacetime, using for instance the version present in \cite{Lehnert}:
\begin{enumerate}
  \item \textbf{Poincar\'e covariance of the Hilbert space where the theory is set.}\\
   This means that unitary operators $U(\Lambda,\,a)$ exist that implement Lorentz transformations and space-time translations;
  \item \textbf{Existence of a vacuum state $|0\rangle$}.\\
   There is a unique state, the vacuum, that is unaffected by Poincar\'e transformations up to a phase: $|U(\Lambda,\,a)|0\rangle=e^{i\phi}|0\rangle$. This implies that this state can only have null four-momentum and angular momentum, since these quantities change under Lorentz transformations. The~vacuum state must have even the lowest allowed energy and must be cyclic, which is acting on it via the creation operators every Hilbert space state can be constructed;
  \item \textbf{Fields constructed via operators.}\\
   All physical quantities can be constructed using polynomials of fields acting on the Hilbert space. The~fields transform under Poincar\'e symmetry as scalars, spinors, and tensors. The~fields are defined in such a way that each one corresponds to a definite physical state, i.e., a particle with defined physical quantities and quantum numbers, such as mass, spin etc.;
  \item \textbf{Energy positivity.}\\
  The Hamiltonian operator is supposed to have non negative eigenvalues. This property together with covariance under the action of Lorentz group implies that the physical four-momentum belongs to the light cone;
  \item \textbf{Microscopic causality.}\\
   Causality is imposed in the meaning of locality, that is, the field operators can commute/anticommute only if they are defined on points separated by space-like vectors: $[\phi(x),\,\phi(y)]_{\pm}=0$ for all space-time points such that $(x^{\mu}-y^{\mu})\eta_{\mu\nu}(x^{\nu}-y^{\nu})<0$.
\end{enumerate}

A QFT is a Lorentz invariant if and only if it is covariant both in and out of the light cone. ``In cone'' means that the momenta are physical, i.e., they belong to the forward light cone in momenta space. Analougously ``out of cone'' means that the considered momenta are not physical. Under these axioms, QFT results totally defined by the vacuum expectation values of the product of field operators. These functions are called {Wightman functions} and are defined as:
\begin{equation}
\label{6}
\mathcal{W}(\Delta x_{1},\,\ldots,\Delta x_{n})=\langle0|\psi(x_{0}),\ldots,\psi(x_{n})|0\rangle
\end{equation}
where $\Delta x_{i}=x_{i-1}-x_{i}$. They depend only on the space-time points differences because of the translation invariance postulated in axiom 1 and 3.

\subsection{Complex Lorentz Group}
In this part of the work, illustrating the Jost argumentation, we follow the exposure of \cite{Greenberg}, which is based on the fact that the space-time inversion is included in the connected component of the complexified Lorentz group. Hence it is necessary to review its properties.\\
The real Lorentz group is the real set of the $4\times4$ matrices that preserve the Minkowski metric:
\begin{equation}
\label{7}
\Lambda^{\alpha}_{\,\mu}\eta_{\alpha\beta}\Lambda^{\beta}_{\,\nu}=\eta_{\mu\nu}\;\Rightarrow\;\Lambda^{T}\eta\Lambda=\eta
\end{equation}
that is $SO(1,\,3)$. This group splits in four disconnected components $\mathcal{L}^{\uparrow}_{+}$, $\mathcal{L}^{\downarrow}_{+}$, $\mathcal{L}^{\uparrow}_{-}$, and $\mathcal{L}^{\downarrow}_{-}$ classified according to the sign of the time changing coefficient $\Lambda^{0}_{\,0}$ and the determinant of the matrix.

Since it is not possible to continuously change the determinant of a transformation, the space inversion matrix is not connected to the identity, and from this there is the necessity to employ the complexified Lorentz group, in particular its covering group: $\overline{\mathcal{L}}^{\uparrow}_{+}=SL(2,\,\mathbb{C})$. It is well known that $SL(2,\,\mathbb{C})$ has two fundamental representations, given by the spinors transformation rules:
\begin{equation}
\label{8}
\begin{split}
&x'_{\alpha}=A_{\alpha\beta}x_{\beta}\\
&\dot{x}'_{\dot{\alpha}}=A_{\dot{\alpha}\dot{\beta}}\dot{x}_{\dot{\beta}}
\end{split}
\end{equation}
where spinors are divided in dotted and undotted. Every $SL(2,\,\mathbb{C})$ representation uses a spinor with $m$ undotted and $n$ dotted indices. The~covering group of the complex Lorentz $\mathcal{L}_{+}$ subgroup is given by the direct product $SL(2,\,\mathbb{C})\otimes SL(2,\,\mathbb{C})$. In this context it is possible to identify a class of transformations that continuously change from the identity to the space-time inversion:
\begin{equation}
\label{9}
\Lambda=\left(
          \begin{array}{cccc}
            \cos{\phi/2} & i\sin{\phi/2} & 0 & 0 \\
            i\sin{\phi/2} & \cos{\phi/2} & 0 & 0 \\
            0 & 0 & \cos{\phi/2} & -\sin{\phi/2} \\
            0 & 0 & \sin{\phi/2} & \cos{\phi/2} \\
          \end{array}
        \right).
\end{equation}

\subsection{Axiomatic Cpt Theorem Demonstration}
In this section we illustrate the core of the Jost argumentation. Considering a $\psi^{(m,\,n)}(x)$ spinor field, it transforms under the action of the Poincar\'e group as:
\begin{equation}
\label{10}
U(\Lambda,\,a)\psi^{(m,\,n)}(x)U(\Lambda,\,a)^{\dagger}=S^{mn}(\Lambda)^{-1}\psi^{(m,\,n)}(\Lambda x)
\end{equation}
where $S^{mn}(\Lambda)^{-1}$ is an opportune matrix determined by the $\Lambda$ transformation. The~generic Wightman function becomes:
\begin{equation}
\label{11}
\begin{split}
&\langle0|\psi^{m_{1}n_{1}}(x_{1}),\ldots,\psi^{m_{k}n_{k}}(x_{k})|0\rangle=\langle0|U(\Lambda,\,a)^{\dagger}U(\Lambda,\,a)\psi^{m_{1}n_{1}}(x_{1})U(\Lambda,\,a)^{\dagger}U(\Lambda,\,a),\ldots,|0\rangle=\\
=&\prod_{i}S^{m_{i}n_{i}}(\Lambda)^{-1}\langle0|\psi^{m_{1}n_{1}}(\Lambda x_{1}),\ldots,\psi^{m_{k}n_{k}}(\Lambda x_{k})|0\rangle
\end{split}.
\end{equation}

Due to the translation invariance, the Wightman function can be written as:
\begin{equation}
\label{12}
\mathcal{W}^{k}(\Lambda\Delta x_{1},\ldots,\Lambda\Delta x_{k})=\prod_{i}S^{m_{i}n_{i}}(\Lambda)^{-1}\mathcal{W}^{k}(\Delta x_{1},\ldots,\Delta x_{k}).
\end{equation}

Now it is necessary to prove the existence of a complex analytic continuation of these functions. One must consider the action of translation on a Wightman function:
\begin{equation}
\label{13}
\begin{split}
&\langle|\psi(x_{1}),\ldots\psi(x_{j})U(\mathbb{I},\,a)\psi(x_{j+1})\ldots|0\rangle=W^{(k)}(\Delta x_{1},\ldots,\Delta x_{j-1},\,\Delta x_{j}+a,\ldots,\,\Delta x_{k})=\\
=&\langle0|\psi(x_{1}),\,\ldots,\,\psi(x_{j}),\,\sum_{p_{r}}p_{r}\rangle\langle p_{r}|e^{-i\,p_{r}a}\psi(x_{j+1}),\ldots,\,\psi(x_{k})|0\rangle
\end{split}
\end{equation}
where the identity expansion $\mathbb{I}=\sum_{p_{r}}p_{r}\rangle\langle p_{r}|$ has been introduced. From the previous relation it follows that the momenta $p_{j}$ and the space-time differences $\Delta x_{j}$ are conjugate coordinates. As a result, one can write a generic Wightman function as a Fourier transform.
\begin{equation}
\label{14}
\mathcal{W}^{k}(\Delta x_{1},\ldots,\,\Delta x_{k})=\frac{1}{(2\pi)^{4(n-1)}}\prod_{i=1}^{k}\int dp_{i}e^{-i\,p_{i}}\,\Delta x_{i}\tilde{W}^{(k)}(p_{1},\ldots,\,p_{k})
\end{equation}
where the $p_{j}$ momenta are physical, that is they belong to the forward light cone, as it is stated by the axiom 4. The~exponential present in this Fourier decomposition must provide a decaying factor in order to guarantee that this distribution behaves as an analytic function. This condition is satisfied when the real part of the exponent is negative: $\mathcal{R}e(-i\,p_{j}\Delta x_{j})<0$. This is possible only if $\mathcal{I}m\Delta x_{j}\neq0$ and in particular $\mathcal{I}m\Delta x_{j}$ is defined on the backward light cone, and this domain is called $T_{k-1}$. Now one can use this result to generalize the Lorentz symmetry. Equation (\ref{12}) remains valid for determined complex valued points $\Delta x_{j}$. Now it is possible to use a theorem by Bargman, Hall, and Wightman \cite{Wightman2}, which states that if Equation (\ref{12}) is still valid, the function $W^{(k)}$ has a defined and unique analytic continuation on the domain $\bigcup \Lambda T_{k-1}=T'_{k-1}$. $T_{k-1}$ is the domain of validity for Equation (\ref{12}) and $\Lambda$ are elements of the complex Lorentz $\mathcal{L}_{+}$ group. The~$\{x_{i}\}$ points belonging to this set are called Jost points and are characterized by the feature that all  sums $\sum_{i}\lambda_{i}(x_{i}-x_{i+1})$ are space-like vectors for every $\lambda_{i}\geq0$ and $\sum_{i}\lambda_{i}>0$. From this it follows that $\sum_{i}\lambda_{i}x_{i}\sim0$ implies $x_{i}\sim0$. Since the complex Lorentz transformations include space-time inversion, from Equation (\ref{12}) one can obtain the relation:
\begin{equation}
\label{15}
\mathcal{W}^{k}(\Delta x_{1},\ldots,\,\Delta x_{k})=(-1)^{L}\mathcal{W}^{k}(-\Delta x_{1},\ldots,\,-\Delta x_{k}).
\end{equation}

It can be shown that the  factor $(-1)^{L}$ is the correct form of the matrix $S^{m_{i}n_{i}}(\Lambda)$ for the space-time inversion. The~next step of the demonstration considers the $k$ points vacuum expectation functions. From the characteristic of fermionic fields, it is possible to demonstrate the following equality:
\begin{equation}
\label{16}
\langle0|\psi^{(n_{1},m_{1})}(x_{1}),\ldots,\,\psi^{(n_{k},m_{k})}(x_{k})|0\rangle=i^{N}\langle|0\psi^{(n_{k},m_{k})}(x_{k}),\ldots,\,\psi^{(n_{1},m_{1})}(x_{1})|0\rangle
\end{equation}
if $N$ fermionic fields are involved. The $N$ coefficient derives from the \emph{spin-statistic theorem, that is from the $T$ transpositions of the $N$ fermionic fields $T=(N+(N-1)+\ldots)=\frac{1}{2}N(N+1)$ with $N$ even. Hence $(-1)^{T}=i^{N}$}. This condition is called \emph{weak local commutativity} at Jost point. Using Equation (\ref{15}) written for the fields:
\begin{equation}
\label{17}
\langle0|\psi^{(n_{1},m_{1})}(x_{1}),\ldots,\,\psi^{(n_{k},m_{k})}(x_{k})|0\rangle=(-1)^{L}\langle0|\psi^{(n_{1},m_{1})}(-x_{1}),\ldots,\,\psi^{(n_{k},m_{k})}(-x_{k})|0\rangle
\end{equation}
together with relation (\ref{16}) one obtains:
\begin{equation}
\label{18}
\langle0|\psi^{(n_{1},m_{1})}(x_{1}),\ldots,\,\psi^{(n_{k},m_{k})}(x_{k})|0\rangle=i^{N}(-1)^{L}\langle0|\psi^{(n_{k},m_{k})}(-x_{k}),\ldots,\,\psi^{(n_{1},m_{1})}(-x_{1})|0\rangle.
\end{equation}

In terms of Wightman functions the previous equality becomes:
\begin{equation}
\label{19}
\mathcal{W}^{k}(\Delta x_{1},\ldots,\,\Delta x_{k-1})=i^{N}(-1)^{L}\mathcal{W}^{k}(-\Delta x_{k},\ldots,\,-\Delta x_{1}).
\end{equation}

This last result is obtained considering {weak local commutativity} with space-time reflection. From Equation (\ref{18}) using the complex conjugation, it is possible to obtain the relation:
\begin{equation}
\label{20}
\langle0|\psi^{(n_{1},m_{1})}(x_{1}),\ldots,\,\psi^{(n_{k},m_{k})}(x_{k})|0\rangle=i^{N}(-1)^{L}\langle0|\psi^{(n_{1},m_{1})\dagger}(-x_{1}),\ldots,\,\psi^{(n_{1},m_{1})\dagger}(-x_{1})|0\rangle^{*}
\end{equation}
where $N=\sum_{i}N_{i}$ and:
\begin{equation}
\label{21}
\theta\psi^{(n,m)}(x)\theta^{\dagger}=(-1)^{l}(\pm i)^{N_{i}}\psi^{(n,m)\dagger}(-x)
\end{equation}
is the generic transformation rule for fermionic fields.

From the equality:
\begin{equation}
\label{22}
\langle0|\psi^{(n_{1},m_{1})}(x_{1}),\ldots,\,\psi^{(n_{k},m_{k})}(x_{k})|0\rangle=\langle0|\theta^{\dagger}\theta\psi^{(n_{1},m_{1})}(x_{1})\theta^{\dagger}\theta,\ldots,\,\theta^{\dagger}\theta\psi^{(n_{k},m_{k})}(x_{k})\theta^{\dagger}\theta|0\rangle
\end{equation}
and the vacuum invariance under $CPT$, the theorem follows. Therefore {weak local commutativity} is a necessary and sufficient condition together with Lorentz invariance to guarantee  $CPT$ symmetry preservation.

A proof based on a more complete mathematical formulation can be found in \cite{Greaves}.

\subsection{Lagrangian Field Theory Cpt Theorem Demonstration}
This proof is based on the argumentation of Bell \cite{Bell}, L\``uders \cite{Lueders}, Pauli \cite{Pauli}, and Schwinger \cite{Schwinger,Schwinger1} and follows the review \cite{Lehnert}.

The Lagrangian approach is based on constructive method: It is shown that all the physical meaningful Lagrangian terms must be $CPT$ even. We will give a review of this result for the most significative fields:
\begin{itemize}
  \item Scalar:
  \begin{equation}
  \label{23}
  \phi_{CPT}(x)=\theta\phi(x)\theta^{\dagger}=\phi^{\dagger}(-x);
  \end{equation}
  \item Fermionic field with spin $=\frac{1}{2}$:
  \begin{equation}
  \label{24}
  \psi_{CPT}(x)=\theta\psi(x)\theta^{\dagger}=-\gamma_{5}\psi^{\dagger T}(-x);
  \end{equation}
  \item Bosonic fields with spin $=1$
  \begin{equation}
  \label{25}
  A_{\mu CPT}(x)=\theta A_{\mu}(x)\theta^{\dagger}=-A_{\mu}^{\dagger}(-x).
  \end{equation}
\end{itemize}

In  previous relations, the coordinate inversion follows from $P$ and $T$. The~operator $C$ instead acts on the fields, causing the hermitian conjugation.\\ The four-gradient under the $\theta=CPT$ action remains unchanged:
\begin{equation}
\label{26}
\partial_{\mu CPT}=\theta\partial_{\mu}\theta^{\dagger}=\partial_{\mu}.
\end{equation}

For a generalized tensor field $T^{\mu_{1}\ldots\mu_{n}}(x)$, the $CPT$ transform is:
\begin{equation}
\label{27}
T^{\mu_{1}\ldots\mu_{n}}_{CPT}(x)=\theta T^{\mu_{1}\ldots\mu_{n}}(x)\theta^{\dagger}=(-1)^{n}T^{\mu_{1}\ldots\mu_{n}\dagger}(-x)
\end{equation}
as it can be shown for the Faraday tensor field $F_{\mu\nu}=\partial_{\mu}A_{\nu}(x)-\partial_{\nu}A_{\mu}(x)$ using the previous result~(\ref{25}):
\begin{equation}
\label{27a}
\begin{split}
&F_{\mu\nu\,CPT}=\theta F_{\mu\nu}\theta^{\dagger}=\theta\partial_{\mu}\theta^{\dagger}\theta A_{\nu}(x)\theta^{\dagger}-\theta\partial_{\nu}\theta^{\dagger}\theta A_{\mu}(x)\theta^{\dagger}=\\
=&\partial_{\mu}(-A_{\nu}(-x))-\partial_{\nu}(-A_{\mu}(-x))=\partial_{\mu}A_{\nu}(x)-\partial_{\nu}A_{\mu}(x)=F_{\mu\nu}(x).
\end{split}
\end{equation}

Now to proceed with the proof one has to notice that the Lagrangian density must be a Lorentz scalar, so all the spinorial indices must be contracted. This means that spinors must be always paired to form bilinears: $\overline{\psi}\psi$, $\overline{\psi}\gamma^{\mu}\psi$, $\overline{\psi}\sigma^{\mu\nu}\psi$, etc.

For instance, in the case of a spinor bilinear $\overline{\chi}\psi$, the $\theta=CPT$ action gives:
\begin{equation}
\begin{split}
\label{28}
&(\overline{\chi}(x)\psi(x))_{CPT}=\theta\overline{\chi}(x)\theta^{\dagger}\theta\psi(x)\theta^{\dagger}=\chi^{T}(-x)\gamma_{5}\gamma_{0}^{*}\gamma_{5}\psi^{\dagger T}(-x)=\\
=&-\overline{\chi}^{T}(-x) \gamma_{0}^{*}\psi^{\dagger T\dagger}(-x)=
-\left[-\psi^{T}(-x)\gamma_{0}^{T}\chi^{T\dagger}(-x)\right]^{\dagger}=\\
=&\left[\left(-\chi^{\dagger}(-x)\gamma_{0}\psi^{T}(-x)\right)^{T}\right]^{\dagger}=\left[\overline\chi(-x)\psi(-x)\right]^{\dagger}
\end{split}
\end{equation}
where the anticommutation of spinor fields and of $\gamma_{0}$ and $\gamma_{5}$ matrices has been used and the transposition on the last equality can be neglected since it is applied to a scalar. The~last equality shows that the general $CPT$ transformation rule for spinor bilinears is given by (\ref{23}). With a simple generalization the following relation follows for generic Lorentz invariant Lagrangian densities:
\begin{equation}
\label{29}
\mathcal{L}_{CPT}(x)=\theta\mathcal{L}(x)\theta^{\dagger}=(-1)^{2k}\mathcal{L}^{\dagger}(-x)=\mathcal{L}^{\dagger}(-x)
\end{equation}
where the contraction of Lorentz indices and  even number of fermionic fields has been used. A~physics meaningful Lagrangian must be not only Lorentz invariant, but even hermitian $\mathcal{L}(x)=\mathcal{L}^{\dagger}(x)$. This last property is motivated by the necessity to guarantee the theory of unitarity. This feature used together with Equation (\ref{29}) justifies the relation:
\begin{equation}
\label{30}
\mathcal{L}_{CPT}(x)=\theta\mathcal{L}(x)\theta^{\dagger}=\mathcal{L}^{\dagger}(-x)=\mathcal{L}(-x).
\end{equation}

Finally this last result, with the assumption that the fields interact in the same space-time point leads to the relation for the action:
\begin{equation}
\label{31}
S_{CPT}=\theta S\theta^{\dagger}=\theta\int_{\mathbb{R}^{4}}d^{4}x\,\mathcal{L}(x)\theta^{\dagger}=\int_{\mathbb{R}^{4}}d^{4}x\,\mathcal{L}(-x)=\int_{\mathbb{R}^{4}}d^4z\mathcal{L}(z)=S
\end{equation}
hence the theorem assertion is proved.

\section{CPT Violation Implies Lorentz Invariance Violation}
In this section we resume the Greenberg argumentation \cite{Greenberg2}, which states that a $CPT$ odd QFT necessarily also violates the Lorentz covariance. The~proof uses the previously introduced algebraic QFT principles.

First it is necessary to define the general $\tau$ function as:
\begin{equation}
\label{32}
\tau^{k}(x_{1},\ldots,\,x_{k})=\sum_{\sigma}\Theta(x^{0}_{\sigma(1)},\ldots,\,x^{0}_{\sigma(k)})\mathcal{W}^{k}(x_{\sigma(1)},\ldots,\,x_{\sigma(k)})
\end{equation}
where $\sigma$ is a permutation of a $k$ indices set and $\Theta$ is the time ordering operator. The~proof is based on the idea that if $CPT$ symmetry is violated for every Wightman function, than the related $\tau$ functions and therefore even the complete QFT theory violate Lorentz covariance. The~demonstration proceeds using the idea that if the $\tau$ functions are covariant, the time reversing Lorentz transformation that leaves invariant the Wightman functions must not affect even the $\tau$ functions. This statement must be true even for space-like points. Therefore there must be a complex Lorentz transformation (time inversion) that does not change the Wightman functions but makes negative the time differences between successive points. From the $\tau$ invariance it follows that Wightman functions are equal if their fields order is totally inverted, as stated before. Hence to preserve the $\tau$ functions covariance it is necessary to have {weak local commutativity} of the Wightman functions, a sufficient and necessary condition for the $CPT$ invariance and this demonstrates the assertion. However the idea that CPT violation automatically implies LIV was confuted in \cite{Discussion} and the argument has been widely debated in the  literature~\cite{Discussion2,Discussion3,Discussion4,Discussion5}.

\section{Consequences of CPT Symmetry}
In this section we briefly analyze the consequences of  $CPT$ symmetry, focusing in particular on the relation between particles and antiparticles.

The first part of this section is devoted to the properties of the general physical eigenstates. It is not necessary to deal with the exact form of these states, but only to consider the Lorentz invariant quantum numbers. The~quantum numbers are supposed to arise from continuous symmetries, which are associated to the conserved currents by the Noether theorem. The~conserved currents are composed of the product of fields and fields derivatives and can be used to form hermitian operators whose functional form is space-time independent. Using the property of $CPT$ symmetry of a generic $n$ indices tensor, it is easy to obtain the following relations for a conserved current:
\begin{equation}
\label{33}
j^{\mu}_{CPT}(x)=\theta j^{\mu}\theta^{\dagger}=-j^{\mu}(-x).
\end{equation}

Considering the charge:
\begin{equation}
\label{34}
Q=\int d^{3}x\,j^{0}(t,\,\vec{x})
\end{equation}
it follows:
\begin{equation}
\label{35}
Q_{CPT}=\theta Q\theta^{\dagger}=-\int_{\mathbb{R}^{3}}d^{3}x\,j^{0}(-t,\,-\vec{x})=-\int_{\mathbb{R}^{3}}d^{3}x\,j^{0}(-t,\,\vec{x})=-Q.
\end{equation}

The energy-momentum tensor transforms under $CPT$ as:
\begin{equation}
\label{36}
T^{\mu\nu}_{CPT}(x)=\theta T^{\mu\nu}(x)\theta^{\dagger}=T^{\mu\nu}(-x).
\end{equation}

From this relation and from the definition of momentum:
\begin{equation}
\label{37}
P^{\mu}=\int_{\mathbb{R}^{3}}d^{3}x\,T^{0\mu}(t,\,\vec{x})
\end{equation}
it is possible to obtain:
\begin{equation}
\label{38}
P^{\mu}_{CPT}=\theta P^{\mu}\theta^{\dagger}=P^{\mu}.
\end{equation}

In the same way, starting from the $CPT$ transformation relation valid for the angular \mbox{momentum tensor}:
\begin{equation}
\label{39}
J^{\mu\nu\alpha}_{CPT}(x)=\theta J^{\mu\nu\alpha}(x)\theta^{\dagger}=-J^{\mu\nu\alpha}(-x).
\end{equation}
 The Lorentz group generator is assigned:
\begin{equation}
\label{40}
M^{\mu\nu}=\int_{\mathbb{R}^{3}}d^3x\,J^{0\mu\nu}(t,\,\vec{x})
\end{equation}
it is possible to obtain:
\begin{equation}
\label{41}
M^{\mu\nu}_{CPT}=\theta M^{\mu\nu}\theta^{\dagger}=-M^{\mu\nu}.
\end{equation}

Considering now a generic one-particle state $|p^{\mu},\,j,\,j_{p},\,q\rangle$, where $p$ represents the four-momentum, $j$~is the total angular momentum, $j_{p}$ is the angular momentum projection along the momentum direction, and $q$ is the particle charge, it is possible to obtain:
\begin{equation}
\label{42}
|p^{\mu},\,j,\,j_{p},\,q\rangle_{CPT}=\theta|p^{\mu},\,j,\,j_{p},\,q\rangle=|p^{\mu},\,j,\,-j_{p},\,-q\rangle.
\end{equation}

The $CPT$ transformations for $p^{\mu}$ and $q$ were obtained before. The~total angular momentum is given by the sum of angular momentum and spin $J^{\mu}=L^{\mu}+S^{\mu}$. The~total angular momentum is represented by the three indices tensor $J^{\mu\nu\alpha}(x)$, so the $j_{p}$ projection transforms as $\theta j_{p}\theta^{\dagger}=-j_{p}$. On the contrary, the total angular momentum is not affected by the $CPT$ action since it is associated to the $J^{2}$ squared operator, that is invariant.

Now we give a brief review of the effects of $CPT$ on coupling constants present in the Hamiltonian. There are a lot of issues in defining couplings, but regardless of these ambiguities, one can for instance consider the interaction Hamiltonian:
\begin{equation}
\label{43}
\mathcal{H}_{int}(x)=\lambda A(x)+\overline{\lambda}A_{CPT}(x).
\end{equation}
If $A_{CPT}(x)\neq A(x)$ then the coupling constants $\lambda$ and $\overline{\lambda}$ may differ. $A_{CPT}(x)=\theta A(x)\theta^{\dagger}$ is the $CPT$ conjugate and the $A$ operator is supposed to not contribute to the anti-particle matrix element and viceversa:
\begin{equation}
\label{44}
\begin{split}
&\langle\psi_{out}|\mathcal{H}_{int}|\psi_{in}\rangle=\lambda\langle\psi_{out}|A|\psi_{in}\rangle\\
&\langle\overline{\psi}_{out}|\mathcal{H}_{int}|\overline{\psi}_{in}\rangle=\overline{\lambda}\langle\overline{\psi}_{out}|A_{CPT}|\overline{\psi}_{in}\rangle
\end{split}
\end{equation}

The interaction Hamiltonian transforms therefore under $\theta=CPT$ as:
\begin{equation}
\label{45}
\theta\mathcal{H}_{int}\theta^{\dagger}=\theta(\lambda A+\overline{\lambda}A_{CPT})\theta^{\dagger}=\lambda A_{CPT}+\overline{\lambda}A.
\end{equation}

The generalized Hamiltonian, which must be $CPT$ even, assumes the explicit form:
\begin{equation}
\label{46}
\mathcal{H}(x)=\mathcal{H}_{0}+\mathcal{H}_{int}=\mathcal{H}_{0}+\lambda A(x)+\overline{\lambda}A_{CPT}(x)
\end{equation}
with $\mathcal{H}_{0\,CPT}=\mathcal{H}_{0}$. The~$CPT$ the invariance of $\mathcal{H}$ implying therefore that $\lambda=\overline{\lambda}$, so the coupling constants for particles and antiparticles must be equal.

Another observation can be made on masses, indeed from the dispersion relation:
\begin{equation}
\label{47}
m=\sqrt{p_{\mu}p^{\mu}}
\end{equation}
and from the conjugation property of $p^{\mu}$ it results that every particle must have the same mass of its related antiparticle. For the same reason a particle and its antiparticle must have the same dispersion relation in every $CPT$ even theory.

To conclude this section we underline how it is possible to construct $CPT$ odd terms for an Hamiltonian: It is possible by introducing different masses or dispersion relations or coupling constants for particles and the respective antiparticles. Finally it is always necessary to violate the LI, even if this is only a necessary but not sufficient condition. More extensive analysis can be found in \cite{Lehnert}.

\section{CPT Violation Motivations}
The first motivation to investigate eventual violations of $CPT$ symmetry was given by Hawking~\mbox{\cite{Hawking,Hawking2}}. He argued that gravitational effects can influence the predictability of future and with this the unitarity of a theory. Indeed black holes with their event horizons can limit the observer knowledge of physics. This argument evolved in the {space-time foam} idea \cite{Wheeler,Mavromatos}, where a pure state evolves in a mixed state because of the quantum background effects causing decoherence. This foam is supposed to be made of fluctuations of the spacetime that can manifest such as Planck scale black holes. The~$CPT$ violations can manifest as unitarity violations, since part of the physical information can disappear inside these black holes event horizons. Hence even in the low energy limit and in the flat space-time, approximation gravitational effects can spoil the validity of  $CPT$ symmetry. Defining the $S$ scattering matrix as the operator thank whom an {in} state evolves to an asymptotic {out}~state:
\begin{equation}
\label{48}
|\psi_{out}\rangle=S|\psi_{in}\rangle
\end{equation}
and the density state matrix as:
\begin{equation}
\label{49}
\rho_{in}=|\psi_{in}\rangle\langle\psi_{in}|
\end{equation}
the effects of the quantum foam on the states evolution can induce decoherence and unitarity violation in the following way:
\begin{equation}
\label{50}
\rho_{out}=Tr|\psi_{out}\rangle\langle\psi_{out}|=\tilde{S}|\psi_{in}\rangle\langle\psi_{in}|=\tilde{S}\rho
\end{equation}
where $\tilde{S}$ is a non unitary operator and therefore $\tilde{S}\neq SS^{\dagger}$. This ensures a violation of  strong $CPT$ invariance intended as the standard formulation of this symmetry. However it is not possible to exclude the validity of the weak version of this symmetry. In this last case, $CPT$ is preserved in the state probabilities of the asymptotic states obtained from the evolution of pure initial quantum states. This hypothesis can be supported by ideas muted by black hole thermodynamics. From black hole evaporation the disappeared information can be re-obtained and  unitarity can be restored. A more detailed discussion can be found in \cite{Mavromatos}.

$CPT$ violation is also predicted in string theory, since this symmetry is valid not only for the equation of motion solutions, but even the strings boundary conditions are invariant under the action of the $\theta=CPT$ operator. This statement can be true only if  the vacuum results $CPT$ invariant is even. However this invariance can be spoiled by the non null vacuum expectation value of some operators as foreseen in some string scenarios \cite{Kostelecky,Kostelecky2,Kostelecky3}.

\section{CPT Theorem in Curved Spacetime}
As already shown, the algebraic demonstration of the $CPT$ theorem resorts to the Wightman axioms, which are strictly related to the QFT formulation in Minkowski flat spacetime and cannot be straightforwardly generalized to curved spaces. Generally speaking, a curved spacetime for instance does not posses Lorentz/Poincar\'e symmetry.

A possible way to obtain a generalization of the theorem for curved spacetime consists in axiomatically defining the QFT using  Ordered Product Expansion (OPE) \cite{Hollands,Hollands2}. This construction can be made in a globally hyperbolic spacetime $(M,\,g_{\mu\nu})$ with an orientation defined via an external form $e$ assigned on $M$ and an equivalence class of time functions $f:\,M\longrightarrow\mathbb{R}$, used to introduce time orientation. The~observables of the theory are introduced as elements of a $C\ast$ algebra $\mathcal{A}(M)$ and the allowed states are defined as elements of the set $S(M)$ of the positive linear maps defined on $\mathcal{A}(M)$.

The OPE is introduced giving the collection of the series coefficients:
\begin{equation}
\label{61}
C(M)=\{C^{i_{1},\ldots, i_{n}}_{(j)}(x_{1},\ldots, x_{n},\,y)\, :i_{1},\ldots, i_{n},\,j\in I,\;n\in \mathbb{N}\}
\end{equation}
where $I$ is an index set suitably defined for this case.

The OPE coefficients must satisfy the following axiomatic requirements:
\begin{enumerate}
  \item \textbf{Locality and covariance.}\\
  The indices must transform under the action of a generic $\rho$ transformation as:
  \begin{equation}
  \label{62}
  (\rho^{*}\times\ldots\rho^{\ast}\times\rho^{-1}_{\ast})C^{(i'_{1},\ldots, i'_{n})}_{(j')}[M']=C^{(i_{1},\ldots, i_{n})}_{(j)}[M]
  \end{equation}
  where $M'$ is the transformed manifold;
  \item \textbf{Identity element..}\\
  \begin{equation}
  \begin{split}
  \label{63}
  &C^{i_{1},\ldots\mathbf{i},\ldots, i_{n}}_{(j)}(x_{1},\ldots, x_{n},\,y)=\\&=C^{i_{1},\ldots, i_{k-1},\,i_{k+1},\ldots, i_{n}}_{(j)}(x_{1},\ldots x_{k-1},\,x_{k+1},\ldots, x_{n},\,y)
  \end{split}
  \end{equation}
  where $\mathbf{i}$ stands for the identity $\mathbb{I}$ in the $k$-place;
  \item \textbf{Compatibility with $\ast$}\\
  \begin{equation}
  \label{64}
  C^{i_{1},\ldots, i_{n}}_{(j)}(x_{1},\ldots, x_{n},\,y)\sim C^{i^{\ast}_{1},\ldots, i^{\ast}_{n}}_{(j^{\ast})}\pi_{0}(x_{1},\ldots, x_{n},\,y)
  \end{equation}
  for an opportune $\pi_{0}$ permutation;
  \item \textbf{Commutativity-anticommutativity.}
  \begin{equation}
  \begin{split}
  \label{65}
  &C^{i_{1},\ldots, i_{k+1},\,i_{k},\ldots, i_{n}}_{(j)}(x_{1},\ldots, x_{k+1},\,x_{k},\ldots, x_{n},\,y)=\\&=-(-1)^{F(i_{k})F(i_{k+1})}C^{i_{1},\ldots, i_{n}}_{(j)}(x_{1},\ldots, x_{n},\,y)
  \end{split}
  \end{equation}
  with $F(i_{k})$ and $F(i_{k+1})$ indices related to the bosonic or fermionic nature of the field involved;
  \item \textbf{Scaling degree.}\\
  \begin{equation}
  \label{66}
  sd\{C^{(i)(j)}_{(k)}\}\leq dim(i)+dim(j)-dim(k);
  \end{equation}
  \item \textbf{Asymptotic positivity.}\\
  $dim(i)\geq 0$ and $dim(i)=0$ if and only if $i=\mathbb{I}$;
  \item \textbf{Spectrum condition.}\\
  The singularities on the field product must have a positive frequency;
  \item \textbf{Associativity.}\\
  An opportunely defined notion of associativity is required;
  \item \textbf{Analytic dependence upon the metric.}\\
  The OPE coefficients must be regular functionals of the space-time metric.
\end{enumerate}

In this work we avoid discussing in more detail  mathematical issues that can arise from  previous requirements. We  limit the discussion to underline that the Wightman functions can be written as:
\begin{equation}
\label{67}
\langle\phi^{i_{1}}(x_{1}),\ldots\phi^{i_{n}}(x_{n})\rangle=\sum_{j}C^{(i_{1},\ldots i_{n})}_{j}(x_{1},\ldots x_{n})\langle\phi^{j}(y)\rangle
\end{equation}
and resorting to the OPE an axiomatic construction of a QFT can be obtained.

\subsection{Construction of a QFT from OPE Coefficients}
As previously mentioned, by using the OPE coefficients collection one can axiomatically define a QFT even in a curved spacetime. The~QFT is given by the pair $\{\mathcal{A}(M),\,S(M)\}$ where $\mathcal{A}(M)$ is a $\ast$-algebra and $S(M)$ is the space of states assigned on $\mathcal{A}(M)$ and defined via the OPE coefficients for a determined $M$ space-time structure. The~algebra is constructed factoring the $\ast$-algebra generated by expressions of the form $\phi^{(i)}(f)$ with $f$ a compactly supported test function associated with the tensorial or spinorial character of $\phi^{(i)}$. The~factorization is constructed satisfying the axiomatic relations:
\begin{enumerate}
   \item \textbf{Linearity.}\\
   $\forall a_{i}$ and $\forall f_{i}$ must be true that:
   \begin{equation}
   \label{68}
   \phi^{i}(\sum_{j}a_{j}f_{j})=\sum_{j}a_{j}\phi(f_{j});
   \end{equation}
   \item \textbf{Existence of $\ast$ operator.}\\
   \begin{equation}
   \label{69}
   [\phi^{(i)}(f)]^{\ast}=\phi^{i\ast}(\overline{f});
   \end{equation}

   \item \textbf{Relations arising from OPE.}\\
   If $O(\cdot)$ is a smeared quantum field, then $O(y)$ is a well defined algebra element for every point $y$, that is if $\langle O(y)|O(y)^{\ast}\rangle=0$ then $O(y)=0$ as an algebra element;
   \item \textbf{Anticommutation Relation.}
   \begin{equation}
   \label{70}
   \phi^{i_{1}}(f_{1})\phi^{i_{1}}(f_{2})=(-1)^{K(i_{1}i_{2})}\phi^{i_{1}}(f_{2})\phi^{i_{1}}(f_{1})
   \end{equation}
   for a properly defined index $K(i_{1}i_{2})$;
   \item \textbf{Positivity.}\\
   \begin{equation}
   \label{71}
   \langle A^{\ast}A\rangle\geq0 \qquad \forall A\in\mathcal{A}(M);
   \end{equation}
   \item \textbf{OPE series expansion.}\\
   \begin{equation}
   \label{72}
   \langle\phi^{i_{1}}(x_{1}),\ldots,\phi^{i_{n}}(x_{n})\rangle\sim\sum_{j}C^{(i_{1},\ldots, i_{n})}_{j}(x_{1},\ldots x_{n},\,y)\langle\phi^{j}(y)\rangle;
   \end{equation}
   \item \textbf{Spectrum condition.}\\
   The previously cited \emph{spectrum condition} can be written in the following form:
   \begin{equation}
   \label{73}
   WF\left(\langle\phi^{i_{1}}(x_{1},\ldots\phi^{i_{n}}(x_{n})\rangle\right)\subset\Gamma_{n}(M)
   \end{equation}
   where $WF$ stands for {wave front} and $\Gamma_{n}(M)$ is a properly defined set of points belonging to $M$.
\end{enumerate}
In this QFT scenario, the following lemma is true:

\textbf{Lemma}\\
\emph{The map $\alpha_{L}:\mathcal{A}(M)\longrightarrow\mathcal{A}(M')$ is a $\ast$-homomorphism if:
\begin{equation}
\label{74}
\alpha_{L}:\phi^{(i)}_{M}(f)\longrightarrow\phi^{L'}_{M'}(\psi^{(i)}f)
\end{equation}
where $\psi^{(i)}$ is an embedding isometry that preserves causality, spin structure, and orientation.}
\\

Resorting to this lemma, one can prove the $CPT$ theorem generalized for curved spacetime, which can be formulated in the following way:

\textbf{Theorem}\\
\emph{Defining the space-time structure $\mathbf{M}=(M,\,g_{\mu\nu},\,e,\,T)$ with $dim(M)=2n$ and defining\\ $\overline{\mathbf{M}}=(M,\,g_{\mu\nu},\,e,\,-T)$, with $\theta^{CPT}_{M}:\mathcal{A}(M)\longrightarrow\mathcal{A}(M)$, one obtains for even $n$ values:
\begin{equation}
\label{75}
\theta^{CPT}_{M}:\phi^{(i)}_{M}(f)\longrightarrow\phi^{(i)}_{\overline{M}}(f)^{\ast}\cdot i^{F(i)}(-1)^{U(i)}
\end{equation}
and for odd $n$:
\begin{equation}
\label{76}
\theta^{CPT}_{M}:\phi^{(i)}_{M}(f)\longrightarrow\phi^{(i)}_{\overline{M}}(f)^{\ast}\cdot i^{F(i)+U(i)-P(i)}
\end{equation}
where $F(i)$ is the number of spinor indices and $F(i)=U(i)+P(i)\;mod\,2$ with $U(i)$ the unprimed (that is the unchanged) and $P(i)$ the primed (that is the transformed) indices, and $\theta^{CPT}_{M}$ is the antilinear isomorfism already introduced.}
\\

The key idea of the demonstration consists in choosing the function $\psi_{(i)}$ of the previous {Lemma} equal to the multiplication of the antilinear immersion: $id:\,V_{M}(i)\longrightarrow V_{\overline{M}}(i\ast)$ by the coefficient $i^{F(i)}(-1)^{U(i)}$ when $n$ is even or by the coefficient $i^{F(i)+U(i)-P(i)}$ if the coefficient $n$ is odd. It can be demonstrated that the $\ast$-homomorphism constructed in this way is an isomorphism proving the~assertion.

\section{CPT and Gravity}
Since the discovery of antimatter, many physicists have wondered about  antimatter behavior in a gravitational field. Many theoretical arguments sustain that particles and antiparticles have the same gravitational properties. Indeed $CPT$ symmetry affects a particle's internal quantum numbers but presumably leaves unaffected the gravity sector. This statement is supported by the fact that the motion in a gravitational field seems to be symmetric under the time reflection, so thanks to the crossing symmetry of QFT a particle falling follows the same geodesic as for an antiparticle motion, with reversed time. Obviously  gravitational behavior can be more complex if the mass is different for particles and antiparticles. What can happen is that antimatter is self-attractive as matter, but the gravitational mutual interaction may be either attractive or repulsive. Arguments against any gravitational repulsive interactions are present \mbox{in \cite{Morrison,Schiff,Good}}, but were questioned and criticized in~\cite{Nieto,Chardin,Chardin2,Hajdukovic}, so this issue is still under debate. In~some proposed interpretations of antimatter gravity, the gravitational mass of antimatter is supposed with negative charge, meaning antimatter is self attractive but interacts repulsively with ordinary \mbox{matter \cite{Hajdukovic,Hajdukovic2}}. Instead other authors have argued that antimatter can be self repulsive \cite{Noyes,BenoitLevy}. The~issue arises from the interpretation of the {charge} present in the formulation of General Relativity (GR) equations of motion. Here we discuss the idea presented in \cite{Villata}, based on the consideration that in GR, the real charge is not the gravitational mass $m_{g}$, but the four momentum $p^{\mu}=m_{g}\frac{dx^{\mu}}{dt}$. The~gravitational masses for matter and antimatter are supposed as positive, together with the related energies and GR is supposed invariant under the action of $CPT$. $CPT$ theorem for curved spacetime has not yet been demonstrated, but in the previous section we have given an argumentation on how it can be extended to curved spacetime under certain conditions. Supposing the validity of the $CPT$ theorem in GR, one can notice that the Einstein field equation is written using 2-indices tensors:
\begin{equation}
\label{51}
R_{\mu\nu}-\frac{1}{2}R\,g_{\mu\nu}=8\pi\,G\,T_{\mu\nu}.
\end{equation}

The Einstein equation results therefore invariant under the action of $CPT$, as it follows from the transformation rule for even-indices tensor fields. Antimatter behaves therefore like ordinary matter and must be gravitationally self attractive. This result can be shown even more directly resorting to the geodesic equation of motion:
\begin{equation}
\label{52}
m_{i}\frac{d^2x^{\mu}}{d\tau^2}=-m_{g}\frac{dx^{\alpha}}{d\tau}\Gamma_{\alpha\beta}^{\mu}\frac{dx^{\beta}}{d\tau}
\end{equation}
where the inertial mass $m_{i}$ and  gravitational mass $m_{g}$ are explicitly written. From the form of \mbox{Equation (\ref{52})}, it is simple to notice that the charge is given by the term $m_{g}\Gamma_{\alpha\beta}^{\mu}\frac{dx^{\beta}}{d\tau}$. The Christoffel~symbol:
\begin{equation}
\label{53}
\Gamma_{\alpha\beta}^{\mu}=\frac{1}{2}g^{\mu\nu}\left(\partial_{\alpha}g_{\nu\beta}+\partial_{\beta}g_{\nu\alpha}-\partial_{\nu}g_{\alpha\beta}\right)
\end{equation}
is a 3-indices object, so it results to $CPT$ odd. The~differential form $dx^{\mu}$ is $CPT$ odd, therefore the terms:
\begin{equation}
\label{54}
\begin{split}
&\frac{dx^{\mu}}{d\tau}\,\Rightarrow\,\left(\frac{dx^{\mu}}{d\tau}\right)_{CPT}=-\frac{dx^{\mu}}{d\tau}\\
&\frac{d^{2}x^{\mu}}{d\tau^{2}}\,\Rightarrow\,\left(\frac{d^{2}x^{\mu}}{d\tau^{2}}\right)_{CPT}=-\frac{d^{2}x^{\mu}}{d\tau^{2}}
\end{split}
\end{equation}
are odd under $CPT$ action. The~$CPT$ operator applied on the test body acts on the following terms of Equation (\ref{52}):
\begin{equation}
\label{55}
\bigg\{m_{i}\frac{d^2x^{\mu}}{d\tau^2},\;\; \frac{dx^{\alpha}}{d\tau},\;\;\frac{dx^{\beta}}{d\tau}\bigg\}.
\end{equation}

The $CPT$ action on the gravitational field generator transforms the term:
\begin{equation}
\label{56}
\big\{\Gamma_{\alpha\beta}^{\mu}\big\}.
\end{equation}

The action of the $CPT$ operator on the test body and  gravitational field generator leaves unaffected the equation of motion (\ref{52}), so antimatter is self attractive like ordinary matter. It is interesting to notice what happens instead for antimatter in a gravitational field generated by ordinary matter, or vice versa. The~motion of antimatter in an ordinary gravitational field is described by the equation:
\begin{equation}
\begin{split}
\label{58}
&m_{i}\left(\frac{d^2x^{\mu}}{d\tau^2}\right)_{CPT}=-m_{g}\left(\frac{dx^{\alpha}}{d\tau}\right)_{CPT}\Gamma_{\alpha\beta}^{\mu}\left(\frac{dx^{\beta}}{d\tau}\right)_{CPT}\,\Longrightarrow\,m_{i}\left(-\frac{d^2x^{\mu}}{d\tau^2}\right)=\\
&=-m_{g}\left(-\frac{dx^{\alpha}}{d\tau}\right)\Gamma_{\alpha\beta}^{\mu}\left(-\frac{dx^{\beta}}{d\tau}\right).
\end{split}
\end{equation}

The opposite situation, that is the motion of matter in a gravitational field generated by antimatter, is described by:
\begin{equation}
\label{59}
m_{i}\frac{d^2x^{\mu}}{d\tau^2}=-m_{g}\frac{dx^{\alpha}}{d\tau}\left(\Gamma_{\alpha\beta}^{\mu}\right)_{CPT}\frac{dx^{\beta}}{d\tau}\,\Longrightarrow\,m_{i}\frac{d^2x^{\mu}}{d\tau^2}=-m_{g}\frac{dx^{\alpha}}{d\tau}\left(-\Gamma_{\alpha\beta}^{\mu}\right)\frac{dx^{\beta}}{d\tau}.
\end{equation}

In both cases the equation of motion becomes:
\begin{equation}
\label{60}
m_{i}\frac{d^2x^{\mu}}{d\tau^2}=m_{g}\frac{dx^{\alpha}}{d\tau}\Gamma_{\alpha\beta}^{\mu}\frac{dx^{\beta}}{d\tau}
\end{equation}
so the gravitational interaction between matter and antimatter is repulsive.

\section{CPT Violation and LIV Research}
The investigation of LIV and $CPT$ violation raises the question of which QFT could describe these departures from standard physics. The~principal environments where these effects are studied are Effective Fields Theories (EFT) developed in order to study the possible phenomenology introduced by LIV or $CPT$ violation in the SM of elementary particles. These theories share the features of preserving the internal $SU(3)\times SU(2)\times U(1)$ gauge symmetry of the standard physics, together with the renormalizability and  formulation made in a traditional perturbative fashion. The~main idea is therefore that the presumed quantum gravity background effects can emerge as perturbations at very high energies and standard physics is again valid in a lower energetic regime.

\subsection{Very Special Relativity}
The first of these effective theories is the {Very Special Relativity} (VSR) theory of Coleman and Glashow~\cite{Glashow-LIV}. In this framework the supposed effects induced by the quantum background modifies the Maximal Attainable Velocity (MAV) of every particle species. This effect implies a modification of a particle's dispersion relations:
\begin{equation}
\label{50a}
E^{2}=p^{2}(1+\epsilon^{2})+m^{2}
\end{equation}
where $\epsilon$ represents the correction factor for the MAV of the particle species taken into account. This modification implies the new propagator form:
\begin{equation}
\label{50b}
D(p^2)=\frac{i}{(p^{2}-m^{2})A(p^{2})+\epsilon^{2}B(p^{2})}.
\end{equation}

The change introduced in the dispersion relation and in the particle's propagator is obviously Lorentz violating, but $CPT$ even, since there is no differences between the Modified Dispersion Relation (MDR) of the particle and the related antiparticle. This theory is conceived in order to be isotropic, but in a preferred reference frame. This means that the Lorentz symmetry is broken.

A following formulation of VSR \cite{VSR} is given in order to break the Lorentz/Poincar\'e symmetry, reducing the invariance from the action of the complete Lorentz group to the action of a subgroup. This new symmetry group is constructed via the generators $T_{1}=K_{x}+J_{y}$ and $T_{2}=K_{y}-J_{z}$ where $J$ are the rotation generators and $K$ the boost ones. This group is called $T(2)$ and is contained in the Lorentz one and is isomorphic to the group of translations in a fixed plane. The~incorporation of $J_{z}$ generates the euclidean motion group $E(2)$, the addition of $K_{z}$ yields the homotheties group $H(z)$, and the incorporation of both $J_{z}$ and $K_{z}$ leads to the similitude group $SIM(2)$. It is important to underline that the addition of one of the discrete operators $P$, $T$, or $CP$ enlarges the symmetry group to the full Lorentz one. It can be shown that it is possible to formulate the second version of VSR using a deformation of the $SIM(2)$ group: The $DISIM_{b}(2)$ generalization \cite{Pope}. The~implementation of this second formulation of VSR leads to the coupling of fields with fixed background tensors and we refer to the next section for a more complete exposition of this approach.

\subsection{Standard Model Extension}
The second framework used to study LIV and even $CPT$ violation effects is the {Standard Model Extension} (SME) \cite{Colladay-Kostelecky}. This model is based on the idea that in some string theories, particular operators can assume vacuum expectation values different from zero and can therefore  can introduce a breaking of space-time isotropy \cite{Kostelecky}. In this theory, the background effects are studied by supplementing all the possible LIV or $CPT$ odd operators, preserving the gauge symmetry and microcausality with respect to the positive energy and four-momentum conservation law. Moreover, the SME formulation is conceived in such a way so as to guarantee the existence of relativistic Dirac and non-relativistic Schr\''odinger equations, in a low energy limit scenario. The~SME modifications consist of perturbation operators, caused by matter fields present in the Lagrangian coupling with background ad hoc introduced tensors. The tensors' constant non-dynamical nature and  vacuum expectation values that are different from zero break the LI. In~SME, furnished operators  preserve renormalizability thanks to their mass dimension $[d]\leq4$, in what is called the minimal SME formulation. In the extended SME formulation, even non renormalizable operators are taken into account. In both  formulations $CPT$ even and $CPT$ odd terms are introduced, as an example we write the SME amended renormalizable quantum electrodynamics Lagrangian:
\begin{equation}
\label{50c}
\begin{split}
\mathcal{L}^{QED}_{LIV}=&\frac{i}{2}\,\overline{\psi}\,\gamma^{\mu}\,\overleftrightarrow{D}_{\mu}\,\psi-m\,\overline{\psi}\,\psi-\frac{1}{4}\,F^{\mu\nu}F_{\mu\nu}-\frac{1}{2}\,H_{\mu\nu}\,\overline{\psi}\,\sigma^{\mu\nu}\,\psi+\frac{i}{2}\,c_{\mu\nu}\,\overline{\psi}\,\gamma^{\mu}\,\overleftrightarrow{D}^{\nu}\,\psi+\\
+&\frac{i}{2}\,d_{\mu\nu}\,\overline{\psi}\,\gamma_{5}\,\gamma^{\mu}\,\overleftrightarrow{D}^{\nu}\,\psi-a_{\mu}\,\overline{\psi}\,\gamma^{\mu}\,\psi-b_{\mu}\,\overline{\psi}\,\gamma_{5}\,\gamma^{\mu}\,\psi-\frac{1}{4}\,k_{\mu\nu\alpha\beta}\,F^{\mu\nu}F^{\alpha\beta}+\\
+&\frac{i}{2}e_{\mu}\,\overline{\psi}\,\overleftrightarrow{D}^{\mu}\,\psi-\frac{1}{2}\,\overline{\psi}\,\gamma_{5}\,\overleftrightarrow{D}^{\mu}\,\psi+\frac{i}{2}\,g_{\lambda\mu\nu}\,\overline{\psi}\,\sigma^{\lambda\mu}\,\overleftrightarrow{D}^{\nu}\,\psi.
\end{split}
\end{equation}
where the $CPT$ odd terms are those with an odd number of contracted Lorentz indices.

Even in SME, the Lorentz covariance is broken modifying the particles dispersion relations. Nowadays SME appears to be the most complete framework to study LIV and $CPT$ violation, since  its rich phenomenological predictions are capable of testing the validity of these fundamental symmetries. As~a final remark,  the SME framework standard physics is perturbed in every regime and presumed phenomenological effects could be detectable in low-energy experiments. Indeed many of the tightest bounds on SME coefficients are set by low-energy atomic and optic experiments \cite{Kostelecky-Russel}.

\subsection{Theories Preserving Covariance}
Another approach to the study of LIV or $CPT$ violations consists in attempting the construction of complete physical theories. This is the case of {Doubly Special Relativity} (DSR) \cite{AmelinoCamelia,AmelinoCamelia2,AmelinoCamelia3,AmelinoCamelia4}, in which LIV is studied formulating a modification of Special Relativity (SR) that attempts to include in the theory formulation another invariant quantity, {the Planck length}, in addition to the light speed. In this model the momentum space and not the spacetime is supposed to be the fundamental structure describing physics.  The concept of absolute locality results relaxed and different observers feel a personal space-time structure, which presents an energy dependence. Space-time description is constructed by every observer in a local way, losing its universality and becomes, therefore, a personal and auxiliary concept that  emerges from the fundamental momentum space, where the real dynamics takes place. The~dispersion relations result again modified and are implemented in the action formulation:
\begin{equation}
\label{50cc}
S=\int\left(\dot{x}^{\mu}p_{mu}-\lambda\left(M(p)-m^2\right)\right)d\tau
\end{equation}
with the particle's MDR $M(p)$:
\begin{equation}
\label{50ccc}
M(P)=|\vec{p}|^2\left(1-f\left(|\vec{p}|,\,{E}\right)\right)-m^2
\end{equation}
where $f\left(|\vec{p}|,\,{E}\right)$ is the perturbation function introduced by the quantum gravity phenomenology. Moreover to determine the momentum space connection, the interaction processes kinematics are modified. More in detail, a modified composition rule for momenta is defined as:
\begin{equation}
\label{50d}
(\mathbf{p},\,\mathbf{q})\rightarrow(\mathbf{p}\oplus \mathbf{q})=\mathbf{p}+\mathbf{q}+g(\mathbf{p},\,\mathbf{q}),
\end{equation}
where $g(\mathbf{p},\,\mathbf{q})$ represents a perturbation of the usual momenta sum. In an interaction process, a particle is therefore supposed by influencing the other incoming particles proportionally to its energy. In this theory the Lorentz symmetry is modified and not simply broken.

Finally, another scenario where the LIV effects are analyzed is the one denoted as {Homogeneously Modified Special Relativity} (HMSR) \cite{Torri,TorriPhD,Antonelli-Miramonti-Torri}. This theory was developed in order to preserve, in addition to the SME symmetries, and is also a covariant formulation. Indeed in this model the Lorentz symmetry is modified and the Lorentz group results amended in order to preserve covariance. In this last theory the interaction with the background is geometrized, indeed the dispersion relations are modified perturbing the kinematic. This implies a modification of the underlying geometry of SR. This model introduces a peculiar form of MDRs that present a perturbation homogeneous of zero degree:
\begin{equation}
\label{50e}
MDR:F(p^2)=\,E^2-\left(1-f\left(\frac{|\vec{p}|}{E}\right)\right)|\vec{p}|^2=m^2.
\end{equation}

In this way the dispersion relation satisfies the requirements act to be a Finsler pseudo-norm. The~peculiarity of the model is indeed the fact that, due to the particular mathematical choice for the tiny LIV perturbative corrections in the MDR, a metric structure is preserved in Finsler geometry  and the MDR isometries define ``modified'' Lorentz transformations, under which the new re-defined  Mandelstam variables are invariant. From the norm, the metric of the momenta space is computed and a direct correspondence of the momenta and coordinate space in Hamilton/Finsler geometry is obtained via the Legendre transform:
\begin{equation}
\label{50ee}
g_{\mu\nu}(\vec{p},\,E)=\left(
                          \begin{array}{cc}
                            1 & 0 \\
                            0 & (1+f(\vec{p},\,E))\mathbb{I}_{3\times3} \\
                          \end{array}
                        \right).
\end{equation}

The dispersion relations are supposed to be different for every particle species, as in VSR \cite{Glashow-LIV}. The~model built in this way is an extension of the SM, which preserves the internal symmetry $SU(3) \times SU(2) \times U(1)$ and does not introduce any exotic particle or interaction and is $CPT$ even. Moreover the isotropic LIV corrections are conceived in order to obtain a modified Lorentz covariant formulation that implies the preservation of space-time isotropy.

\subsection{CPT Violation and LIV Geometry Framework}
A theoretical model including the possibility of LIV or CPT violation usually also aims  to find an appropriate geometry to formulate an extended version of the ``usual'' theory, making possible some kind of unification between quantum field theory and GR, i.e., QFT and gravity \cite{Kosteleckyfin}. Indeed it is well known that the conventional Riemann geometry, even in the extended Riemann/Cartan formulation is incompatible with Lorentz or $CPT$ violation. The~natural setting of such theories appears to be the Riemann/Finsler geometry \cite{Lammerzahl,Pfeifer1,Bubuianu,Schrek}. This idea has been investigated in the SME framework~\cite{Colladay-Kostelecky}, in which the free particle's kinematic is described by the geodesic motion in a spacetime that emerges due to the coupling of the SME background tensors with the fields describing particles. Finsler geometry is used even in SR modification theories, such as DSR \cite{AmelinoCamelia,AmelinoCamelia2,AmelinoCamelia3,AmelinoCamelia4},  VSR \cite{Glashow-LIV,VSR}, and HMSR \cite{Torri,TorriPhD}, where the Lorentz symmetry is only modified and not merely broken as in SME. In DSR formulation local relativity states that the momentum space is the fundamental structure at the basis of the physical processes description, indeed the momentum space is supposed curved and the underlying geometry appears to be the Finsler one, in which the framework presents an explicit dependence on momenta~\cite{AmelinoCamelia5,Pfeifer}. Even VSR \cite{Fuster} and HMSR \cite{Torri} are settled in the Finsler geometry, which naturally emerges in this case as the appropriate one from the kinematic perturbation introduced via the MDRs \cite{Barcaroli}.

\section{Search for CPT and LIV Violation in Astroparticle Physics}
Since in some theoretical models the supposed quantum gravity effects are expected to be more visible in the high energy limit as LIV perturbations, astroparticle physics can be a useful framework to conduct search on possible departures from $CPT$ symmetry or eventually from the Lorentz covariance. Indeed the tiny effects caused by quantum gravity are supposed to be more visible at the highest energies reached for instance in the astrophysics sector. Moreover these effects are supposed to add up during the particle's propagation for the cosmic distances involved in astroparticle physics. In this section, the presumed LIV effects could manifest as modifications of the photons and the Ultra High Energy Cosmic Rays (UHECRs) propagation and finally LIV can influence the neutrino oscillations.

\subsection{Ultra High Energy Cosmic Rays}
In this section we give a review of the expected effects related to the {Cosmic Rays} (CRs) physics, in particular the enlargement/modification of the Greisen Zatsepin Kutzmin (GZK) opacity sphere \cite{Zatsepin,Greisen} to the propagation of UHECRs, that are CRs with energy equal or bigger than $\sim 5\times 10^{19}\,eV$. The
CRs are massive particles and constitute part of the radiation of extraterrestrial origin.
According to their composition, they are divided in a light component, mainly composed of protons (protons and He bare nuclei), an intermediate one (C, N, and O bare nuclei) and a heavy one composed of an iron-type nuclei (Fe and Ni). During their propagation in free space, CRs interact with the Cosmic Microwave Background (CMB) and  can undergo  physical processes that imply energy dissipation. The~interaction with the CMB is determined by the energy and  nature of the cosmic ray involved. Protons can interact with the CMB through a pair production process, that is dominant for low energies:
\begin{equation}
\label{51e}
p+\gamma\,\rightarrow\,p+e^{-}+e^{+}
\end{equation}
and via a photopion production, that through a $\Delta$ particle resonance is dominant for the high energy~limit:
\begin{equation}
\label{52e}
\begin{split}
&p+\gamma\,\rightarrow\,\Delta\,\rightarrow\,p+\pi^{0}\\
&p+\gamma\,\rightarrow\,\Delta\,\rightarrow\,n+\pi^{+}
\end{split}
\end{equation}
Instead heavy nuclei CR can interact via a photodissocitation process:
\begin{equation}
\label{53e}
A+\gamma\,\rightarrow\,(A-1)+n
\end{equation}
where $A$ is the atomic number of the considered CR bare nucleus.\\
The creation energy of UHECRs is limited by the dimensions of the astrophysical object that accelerated these particles.
As a consequence of this upper limit and of the dissipation mechanism, the UHECRs originated outside a sphere centered on Earth and of a determined radius can be detected only under a determined energetic threshold.
For this reason, free space is revealed to be opaque to the propagation of massive particles such as UHECRs.

There are some experimental hints that the GZK opacity sphere could be modified in respect to  theoretical predictions,
as discussed in the work \cite{Resconi}, where UHECR are correlated with cosmic neutrino sources located farther than the classic opacity length. This modification can be justified by the introduction of LIV perturbations in the standard particle physics. This possibility is investigated for instance in works as \cite{Stecker,Stecker2,Torri2,TorriPhD}, in which the propagation of the light massive cosmic rays (protons and similar particles) is analyzed. The~energy dissipation mechanism caused by the interaction with the CMB determines a modification of the {attenuation length} or the {mean free path} of a proton, defined as the average distance that the particle has to travel in order to reduce its energy by a factor of $\dfrac{1}{e}$.
The~inverse of the attenuation length is given by \cite{Stecker}:
\begin{equation}
\label{54e}
\begin{split}
&\frac{1}{l_{p\gamma}}=\int_{E_{th}}^{+\infty}n(E)\,dE\int_{-1}^{+1}\frac{1}{2}\,s\,(1-v_{p}\,\mu)\,\sigma_{p\gamma}(s)\,K(s)\,d\mu=\\
&\phantom{\frac{1}{l_{p\gamma}}}=\int_{E_{th}}^{+\infty}n(E)\,dE\int_{-1}^{+1}\frac{1}{2}\,s\,(1-v_{p}\cos{\theta})\,\sigma_{p\gamma}(s)\,K(s)\,d\cos{\theta}
\end{split}
\end{equation}
where $\mu$ is the impact parameter $\mu=\cos{\theta}$ and $n(E)$ is the density function of the CMB: A Planck's black body energy distribution, $\sigma_{p\gamma}(E)$ is the cross section for the proton-photon interaction and $E_{th}$ is the threshold energy for this physical process. $K(s)$ is the inelasticity of this reaction, defined as the energy fraction available for the production of secondary particles. One can define the elasticity as the energy fraction preserved by the primary particle after the interaction process, $\eta=\dfrac{E_{out}}{E_{in}}$, where $E_{in}$ is the primary incoming energy and $E_{out}$ is its energy after the interaction. From this relation the definition of the inelasticity $K = 1 - \eta$ follows.

Using the following relations:
\begin{equation}
\label{54ee}
\begin{split}
&s=(m_{p}+\epsilon')^2-\left|\vec{p}_{\gamma}'\right|^2=m_{p}^2+2m_{p}\epsilon'\\
&\epsilon'=\gamma\epsilon(1-v_{p}\cos{\theta})
\end{split}
\end{equation}
where the Mandelstam variable $s$  is computed using the photon four momentum defined in the rest frame of the cosmic ray $(\epsilon',\,\vec{p}_{\gamma}')$. In the high energy limit the proton velocity can be approximated as $v_{p}\simeq1$, with $ds=-2E_{p}\epsilon\,d\cos{\theta}$ and using the explicit form of the $n(E)$ Planck distribution, the previous \mbox{relation (\ref{54e})} can be simplified, obtaining:
\begin{equation}
\label{54f}
\frac{1}{l_{p\gamma}}=-\frac{k_{\text{B}}\,T}{2\,\pi^2\,\gamma^2}\int_{E'_{th}}^{+\infty}E'\,\sigma_{p\gamma}(E')\,K(E')\,\ln{\left(1-e^{-E'/2k_{\text{B}}T\gamma}\right)}\,dE'.
\end{equation}

In the classical Lorentz invariant physics scenario, the inelasticity is given by the relation \cite{Stecker}:
\begin{equation}
\label{54g}
K(s)=\frac{1}{2}\left(1-\frac{m_{p}^{2}-m_{\pi}^{2}}{s}\right).
\end{equation}

The introduction of LIV perturbations in the standard physics determines kinematic modifications, which reduce the phase space part available for the photopion production, amending the inelasticity $K$ (\ref{54g}) in Equation (\ref{54e}). These kinematic modifications determine a reduction of the inelasticity as a function of the proton energy, implying a dilatation of the attenuation length \cite{Stecker,Stecker2,Torri2}.\\
The introduction of LIV can influence even the propagation of heavy CR, modifying the photodissociation process they undergo. For its peculiar feature photodissociation determines a change in the chemical composition of the CR considered, and this means that after every interaction the CR changes. However this process can be described as a dissipation effect grouping the CR in families based on similar chemical composition. Since the photodissociation is a relatively rare process and in the hypothesis  only few nucleons are lost in every process, one can suppose that a UHECR belongs to the same family even after some interactions, so one can study this phenomenon as the proton energy dissipation and can see that even in this case the attenuation length is enlarged \cite{Maccione}.

As already underlined, a $CPT$ odd theory is also a LIV theory, but in the presumed GZK modification the introduction of explicit $CPT$ violation is not directly required.

\subsection{Time Delays}
Among other theoretical and experimental evidences that SR is modified or breaks down in the ultra-high energy scenario, time delay in the propagation of high energetic particles plays an important role.

It is well known that intense gravitational fields can modify the propagation of photons traveling nearby massive objects because of space-time dilation and this effect is a classic test of GR \cite{Shapiro}. However some experimental results hint the possibility that photons propagating with different energies and in the absence of intense gravitational fields can present an analogous time delay.

In this section, the best constraints on the quantum gravity induced effects magnitude can be posed using astrophysical sources with distant, bright and very energetic astrophysical sources such as Gamma-Ray Bursts (GRB), flaring Active Galactic Nuclei (AGN), and Pulsars (PSR). For instance this effect was observed between gamma-ray flares, with different energies, originated from the Markarian 501 galaxy center \cite{Albert}.  This phenomenon can be explained introducing modifications to SR that imply a departure from the Lorentz/Poincar\'e symmetry perturbing the free particle's kinematic both in a Lorentz breaking \cite{Ellis:2018,Abdalla:2019,Xu:2018} or in a SR modification scenario \cite{AmelinoCamelia:1997,AmelinoCamelia:2011,Loret:2014,Aldrovandi}. In the following explanation of this phenomenon, we briefly review the argumentation of the SR kinematic symmetry modification approach.

As shown before modifying the dispersion relations, the resulting geometry acquires an explicit dependence on the particle's momentum, which is propagating particles probe of different spacetimes as a function of their energy. In DSR or desitter projective relativity the line element acquires the explicit~form:
\begin{equation}
\label{55}
ds^2=dt^2-n^2(E)\delta_{ij}dx^{i}dx^{j}
\end{equation}
where $n(E)$ is the function expressing the metric dependence on the particle's energy that is a departure from the standard Minkowski geometry, and is supposedly increasing with the energy, hence it can be written as:
\begin{equation}
\label{55a}
n(E)=\sum_{n=1}^{+\infty}k_{n}\left(\frac{E}{E_{P}}\right)^{n}
\end{equation}
where $E_{P}\simeq10^{19}\,\text{GeV}$ is the Planck energy, which is the characteristic suppression scale for the LIV phenomenon, and $k_{n}$ are positive adimensional coefficients. From the previous relation (\ref{55}), one can compute the photon propagation velocity as a function of its energy, considering that this particle propagates along the null direction:
\begin{equation}
\label{56}
ds^2=0\;\Rightarrow\;dt^2=n^2(E)\delta_{ij}dx^{i}dx^{j}=n^2(E)\,dx^2
\end{equation}
finally one can obtain:
\begin{equation}
\label{57}
|\vec{v}(E)|=\left|\frac{d\vec{x}}{dt}\right|=\frac{1}{n(E)}
\end{equation}
where the metric correction factor $n(E)$ acquires the role of a refractive index and $n(E)>1$. Taking into account that the velocity of a photon depends on its energy, one can obtain the time difference $\Delta t$ between two gamma ray burst with different energies to travel the same distance $L$:
\begin{equation}
\label{58}
\Delta t=\frac{L}{c}\left(n\left(E_{2}\right)-n\left(E_{1}\right)\right)=L\left(n\left(E_{2}\right)-n\left(E_{1}\right)\right)
\end{equation}
where the standard light speed is posed $c=1$. This result is in accordance with the experimental evidence that a higher energy gamma-ray burst is delayed with respect to a lower energetic one, since $n(E_{2})>n(E_{1})$ for $E_{2}>E_{1}$. A complete phenomenological discussion on this topic can be found in~\cite{Bolmont}.

An analogous time delay effect is theoretically predicted even in massive particle's propagation. Indeed starting from a massive particle's MDR written as:
\begin{equation}
\label{59}
E^2=p^2(1-f(\vec{p},\,E))+m^2
\end{equation}
where $f(\vec{p},\,E)$ is the LIV-induced perturbation as a function of the particle's four-momentum. Now it is possible to compute the particle's four-velocity:
\begin{equation}
\label{60}
v^{\mu}=\frac{\partial E}{\partial p_{\mu}}=\frac{p^{\mu}}{\sqrt{p^2(1-f(\vec{p},\,E))+m^2}}-\frac{\partial f(\vec{p},\,E)}{\partial p_{\mu}}\frac{1}{2\sqrt{p^2(1-f(\vec{p},\,E))+m^2}}
\end{equation}
obtaining a result that depends on the particle's momentum.

\subsection{CPT and LIV in Neutrino Physics}
The consequences of presumed LIV on neutrino phenomenology are illustrated, for instance, \mbox{in~\cite{Antonelli-Miramonti-Torri,TorriPhD,Torri3}}. The~idea of introducing LIV in neutrino physics  was first investigated in 1999 by Coleman and Glashow ~\cite{Glashow-LIV}.

The presence of tiny perturbative corrections, breaking Lorentz invariance determines a modification of the dispersion relations, introducing the idea of a MAV (that can differ from one particle to the other) modifying the neutrino propagation and, consequently, the oscillation phenomenon. This aspect is
common to  different studies of LIV, with and without the violation of $CPT$ invariance, which followed over the years.

Some of these works consider also very exotic scenarios that are difficult to reconcile with  present phenomenological data or predicting phenomenogical effects that are difficult to distinguish from the standard scenario.
This is the case, for instance, of the already cited VSR model, in which the ``magnitude'' of Lorentz violating effects is  limited  by the connection between LIV and the violation of some internal symmetry of the theory ($CP$ symmetry in this case), as underlined in~\cite{VSR}. In some of the papers adopting this model~\cite{Glashow-LIV,VSR2}, the authors also consider  a possible ``natural origin'' of neutrino oscillation, different from the usual one, which would require neither the introduction of additional states, nor the lepton number violation, but would imply ultra-relativistic neutrinos ($\gamma > > 1$). In this case, VSR and conventional neutrino masses would be indistinguishable, but VSR effects could be significant near the beta decay endpoint, where neutrinos are non ultra-relativistic~\cite{VSR2}.

The possibility of ultra-luminal neutrinos and the limits on LIV magnitude coming from high energy cosmic neutrino phenomenology has also been investigated  in papers like \cite{Liberati-Scully}, in which the Lorentz violation hypothesis is introduced in a more general quantum gravity framework, following an approach based on EFT formalism.

We have already explained in previous sections that the reference work for this approach is the \mbox{SME~\cite{Colladay-Kostelecky}},
investigating both $CPT$-conserving and $CPT$-violating possible sources of LIV.
The experimental tests aim to put limits on the possible values of  different terms in the SME lagrangian that involve different sectors of elementary particle physics (for what concerns the photon sector a comprehensive analysis of potential experimental testable effects, together with a complete analysis of the possible LIV contributions, in case also $CPT$-symmetry is violated, is performed in \cite{Kosteleckymewes09}), electrons, muons, tauons, nucleons, and quarks),
as summarized in a masterful way in~\cite{Kostelecky-Russel}. With certainty, neutrino phenomenology plays an essential role in such kinds of search.

Coming back to the work of~\cite{Liberati-Scully},
the LIV operators in that case are non-renormalizable and the authors consider both possibilities of $CPT$ conservation and
$CPT$ violation. Significant differences between the two cases emerge by comparing the Monte Carlo simulations with the Ice Cube observed spectrum, considering the 2014 IceCube data~\cite{IceCube-2014}, relative to three years of data taking, which reported the first evidence for a high-energy neutrino flux of extraterrestrial origin (with energies in the range from 30 TeV to 2 PeV). The~authors of~\cite{Liberati-Scully} found that, in the case of $CPT$ even operator dominance, if the drop off in a neutrino flux above $\sim$
2 PeV is caused by Planck scale physics, a pileup effect would be produced just below the drop off energy. On the other hand,
this drop off effect would not be observed if the $CPT$-odd term dominates.

This kind of quantum gravity analysis is based on the idea that we cannot directly access extremely high energies close
to the Planck scale, at which potential corrections due to LIV quantum modifications of the space-time structure would manifest themselves, but, nevertheless, one can in any case look for testable effects at lower energies (accessible to the experiment), that
in an EFT approach would be the ``low energy'' consequences of the very high energy LIV corrections.
The investigation of these testable effects would give information on the signature of the kind of LIV and its magnitude.

In the paper~\cite{Liberati-Scully}, the possible LIV-violating terms (including the non-renormalizable ones) are classified according to their
mass dimensions and the attention is focused on the [d] = 4 and on the lowest order  ([d] = 5 and [d] = 6) Planck-mass suppressed operators and on their effects on extragalactic high energy neutrinos propagation. Similarly to the VST case, the presence of these LIV operators introduces deviations from the standard physics predicted $c$ for every particle's MAV.

For a particle of species $i$ and mass $m$, the MDRs can be rewritten in the form:
\begin{equation}
\label{reldisp}
E^2 - p^2 = \tilde{m}_i^2(E)
\end{equation}
where $\tilde{m}_i$ is a sort of effective mass term, containing the corrections generated by LIV operators: $ \tilde{m}_i^2 = m^2 + \delta_i E^2$. The energy dependent dimensionless (here and in the rest of the paper we adopt, for simplicity, the natural units ${\hbar} = c = 1$) $\delta_i$ coefficients are
functions of the lagrangian parameters, which can be different for different particle species and  determine the MAV for every particle.
The LIV effects would be visible in processes involving at least two different particle species (having different MAV),
$i$~and $j$ (as in the neutrino flavor oscillation case), and the magnitudes
of these effects is determined by the values of the differences $\delta_{ij} = \delta_i -\delta_j$.

The high energy neutrino data
analysis enables to put some constraints on the $\delta_{ij}$ values and consequently on the coefficients of the LIV terms in the
lagrangian. The~$\delta_{ij}$ can be organized as a perturbative expansion in powers of $\left(\dfrac{E}{E_P}\right)^n$ (\ref{55a}), with $E_P$  Planck energy  and $n=[d]-4$:
\begin{equation}
\delta_{ij} = \sum_{n=0,1,2}  k_{\small{i,j,n}} \left(\frac{E}{E_P}\right)^n\, .
\label{eqLIV}.
\end{equation}
The coefficient corresponding to [d] = 4 dimension operator
can be supposed identically zero, or at least suppressed
with respect to the ones of the [d] = 5 and [d] = 6 operators (see e.g, \cite{Liberati-review}). In any case the coefficient of this operator entering in processes in which neutrinos and electrons are involved has been seriously constrained by the analysis \cite{Stecker-Scully-2} of the IceCube data \cite{IceCube-2014}, which fixed $k_{\nu_x,e,0} < 5.2 \times 10^{-21}$.

Concerning the remaining LIV operators, the ones with odd mass dimensions (for which $n$ is an odd number) are also $\cal{CPT}$-odd, while the other operators (for which $n$ is an even number) are associated to terms in the Lagrangian that violate the Lorentz invariance but preserve $CPT$
symmetry, as discussed in \cite{Kosteleckymewes09}, in which all gauge-invariant Lorentz and $CPT$-violating terms with dimensions up to nine in the quadratic Lagrangian density are classified.

As already underlined, one can distinguish two different classes of theories: The ones assuming the violation of $CPT$ and Lorentz symmetry and the other, which predict LIV but are $CPT$-even.
For both classes of models, the possible constraints on the different operator coefficients coming from experiments with fermions have been analyzed
e.g. in~\cite{Kostelecky2013} and, in the specific case of high energy astrophysical neutrino data and $CPT$-conservation, in~\cite{Diaz-Kostelecky-Mewes-HE-neutrinos}.

Using, the notation (adopted also by~\cite{Liberati-Scully}) of  Equation~(\ref{eqLIV}), the class of $CPT$-violating theories contains models for which the coefficient $k_{\nu_x, e, 1}$
dominates the LIV quantity
$\delta_{ij}$
for processes involving neutrinos and electrons and, consequently, determines also the magnitude of the velocity excess for superluminal neutrinos. In this case, the fact that the theory is $CPT$-odd can imply that the dispersion relations differ for neutrinos and antineutrinos. This implies that neutrinos can be superluminal while antineutrinos are subluminal or vice versa and this offers a characteristic phenomenological signature. For~the {$CPT$}-conserving class of models, instead, the leading coefficient in~(\ref{eqLIV}) is $k_{\nu_x,e,2}$.

In principle one should introduce also a helicity dependence in the coefficient of the n = 1  operator, because in this case (due $CPT$-violation) one has to distinguish the LIV contributions associated in the Lagrangian
to terms (allowed in Standard Model extensions) involving right handed neutrinos. However the potential contributions of these terms to electron coefficients are negligible because they are very strongly constrained by experimental data, mainly by Crab nebula
observation~{\cite{Data-Crab,Data-Crab-2,Data-Crab-3,Data-Crab-4,Data-Crab-5}.

The possibility of using neutrinos from astrophysical sources to test the hypothesis of theories predicting LIV and eventually also $CPT$-violation
has been widely exploited, not only in the studies that we just discussed (most of which rely on the investigation of 2014 Ice Cube data~\cite{IceCube-2014}) and that will also continue  in future (concerning these kind of analyses relying mainly on future Ice Cube data, see also \cite{Ice-Cube-LIV-potentiality}),
but also with the use of other sets of data and considering different astrophysical sources.
For instance, the case of the study of the propagation of neutrinos from a Core-Collapse Supernova (looking for variations in time of the order of
a few milliseconds \cite{LIV-Supernova-Elllis}). This includes even the search for a possible LIV-induced spread out of the time profile signal characteristic of
a $\nu_e$ neutronization burst from the next galactic supernova \cite{SN-LIV-Mirizzi-et-al} and also of the LIV and CPT violation tests by the
oscillation pattern observation in experiments measuring atmospheric neutrinos \cite{LIV-CPT-atmo-1, LIV-CPT-atmo-2}.
The most promising opportunity is probably offered by the recently born and increasingly relevant multimessenger astronomy, looking
for very high energy neutrinos emitted by astrophysical objects (like blazars, neutron stars, and so on) together with other electromagnetic
and/or gravitational signals. Examples of this kind are offered by~\cite{LIV-multimessenger-1,LIV-multimessenger-2}.

In the last few years, the investigation of astrophysical neutrinos as probes for potential signals of Lorentz and $CPT$ symmetry violation has been
flanked with analogous studies performed with artificial neutrino sources, produced mainly by accelerators and also partially by reactors.

A first interesting example is given by~\cite{LIV-e-SBL}, in which the authors discuss the possibility of using a generic Short Baseline (SBL) accelerator neutrino experiment to look for a possible anomalous energy dependence
in the spectrum (differing from the one predicted by usual massive neutrino oscillation) and for a dependence on the direction of neutrino propagation. They apply this idea in particular to the LSND data, obtaining a possible non zero
value of the order of $(3 \pm 1) \times 10^{-19}$ GeV, for a combination of LIV coefficients.
A similar study has been performed by the MiniBooNE Collaboration in~\cite{MiniBoone-LIV-CPT} using both the
$\nu_e$ and $\bar{\nu}_e$ appearance data of this experiment and looking for sidereal time variation effects that could be induced by CPT and Lorentz violations. These effects do not seem to be present in most of the data, but one can find a competitive upper limit  of the order of $10^{-20}$ GeV (similar to that derived in~\cite{LIV-e-SBL}), for some
combination of the SME model coefficients, which would eventually contribute to the
$\nu_{\mu} \to \nu_e$ oscillation channel. A search for such sidereal modulation effects in the neutrino interactions
rate has also been performed  more recently~\cite{T2K-near-LIV}, using the T2K on-axis near detector (therefore for a short value of the baseline, even if in the case of a near detector built for a long baseline experiment (LBL)). The analysis gave no evidence
of such a signal.

An even more interesting possibility is offered by the LBL accelerator neutrino and the medium baseline reactor antineutrino experiments, as discussed since a few years ago in~\cite{Diaz-LIV-e-LBL} and in~\cite{Yufeng-LIV-reactor}, respectively for the first and  second category of experiments.
The opportunity of testing LIV and CPT violation with the LBL experiments has become particularly interesting with the new generation of experiments, already running or planned for the very near future, like T2K~\cite{T2K,T2K-CPT}, MINOS~\cite{MINOS, MINOS-CPT-far,MINOS-CPT-teoria}, DUNE~\cite{DUNE,DUNE-CPT-LIV}, and  NO$\nu$A\cite{NOVA}.
The LBL experiments have
already given fundamental contributions to our understanding of neutrino properties, from the determination of the ``atmospheric'' oscillation parameters
$\theta_{32}$ and $\Delta m^2_{32(31)}$, to the indication of $\theta_{31} \neq 0$ (complementary to the studies performed by reactor antineutrino experiments). The~new generation of LBL is designed to attack some of the main open questions of neutrino physics of great impact for all elementary
particle physics and astrophysics, like the search for CP violation in the leptonic sector and the determination of neutrino mass ordering. A very remarkable result has been obtained recently~\cite{T2K-CP} by T2K, which found important constraints on the possible CP violation phase ($\delta_{CP} \neq 0$ and $\delta_{CP} \neq \pi$ at almost 3 $\sigma$), confirming the hints derived already by other LBL and mainly by
NO$\nu$A.}~\cite{Mezzetto-Terranova}.

The LBL accelerator experiments have automatically made the possibility of studying both the neutrino and antineutrino channels and, therefore, they are the ideal framework for $CPT$ studies.
Moreover this new generation of experiments is characterized by high values of the energy E and of the baseline L, and, therefore, they also offer  the opportunity to search for signals of LIV (which is expected to introduce a subleading effect in the oscillation pattern proportional to $L \times E$).
In addition to this, the high intensity of the beams guarantees high statistics, essential for every investigation looking for tiny effects.

A recent interesting example of such kinds of studies is given by~\cite{LBL-LIV-indiani}. The~authors perform their analysis in the already cited framework of SME~\cite{Colladay-Kostelecky}, restricting  their analysis to the possibility of LIV together with $\cal{CPT}$-violation, like in~\cite{Colladay-Kostelecky2,Diaz-Kostelecky-Mewes-2009}.
Following the line of~\cite{Kostelecky-Mewes-2004} and of~\cite{Kostelecky-Mewes-2012}, in which the possible operators appearing in the SME Lagrangian are classified according to their mass dimension (as already discussed
for~\cite{Liberati-Scully}), and keeping into account only renormalizable terms (that is contributions of mass dimension $[d]\leq4$), the LIV part of the neutrino lagrangian density can be written as:
\begin{equation}
{\cal{L}}  = - \frac{1}{2}  \left[p^{\mu} _{\alpha \beta} \bar{\psi}_{\alpha} \gamma_{\mu} \psi_{\beta} +
q^{\mu} _{\alpha \beta} \bar{\psi}_{\alpha} \gamma_5 \gamma_{\mu} \psi_{\beta}
- i r^{\mu \nu}_{\alpha \beta} \bar{\psi}_{\alpha} \gamma_{\mu} \partial_{\nu} \psi_{\beta}
- i s^{\mu \nu}_{\alpha \beta} \bar{\psi}_{\alpha} \gamma_5 \gamma_{\mu} \partial_{\nu} \psi_{\beta} \right] + h.c.
\label{L-LIV-neut}.
\end{equation}

The different contributions can be reorganized, considering the fact that only the left handed part of the neutrino wave functions takes part in the interactions and keeping only the terms that violate
CPT, in addition to LI. In this way the CPT violating and LIV corrections can be parametrized in terms of just two parameters:
$(a_L)^{\mu}_{\alpha \beta} = (p + q)^{\mu}_{\alpha \beta}$ and $(c_L)^{\mu\nu}_{\alpha \beta} = (r + s)^{\mu}_{\alpha \beta}$, which are represented by hermitian matrices in the flavor space. The~choice adopted in~\cite{Kostelecky-Mewes-2004} and pursued also
in ~\cite{LBL-LIV-indiani}  is that of choosing $\mu = \nu = 0$ in order to realize in the simplest way (but not the only possibility) isotropy for what concern space-time rotations in a fixed preferred reference frame with respect to the isotropy preserving coefficient $(c_L)^{\mu\nu}_{\alpha \beta}$. An important note is that the term $(a_L)^{\mu}_{\alpha \beta}$ explicitly violates Lorentz invariance introducing a preferred direction in spacetime. In this way one can write the LIV part of the Hamiltonian using the matrices $a^0_{\alpha \beta}$ and $(c)^{00}_{\alpha \beta}$:
\begin{equation}
H_{LIV}=
\left(
  \begin{array}{ccc}
    a^{0}_{ee} & a^{0}_{e \mu} & a^{0}_{e\tau} \\
    (a^{0}_{e \mu})^{\ast} &  a^{0}_{\mu \mu} & a^{0}_{\mu \tau} \\
    (a^{0}_{e \tau})^{\ast} & (a^{0}_{\mu \tau})^{*} & a^{0}_{\tau \tau} \\
  \end{array}
\right)
-\frac{4}{3}E
\left(
  \begin{array}{ccc}
    c^{00}_{ee} & c^{00}_{e \mu} & c^{00}_{e\tau} \\
    (c^{00}_{e \mu})^{\ast} &  c^{00}_{\mu \mu} & c^{00}_{\mu \tau} \\
    (c^{00}_{e \tau})^{\ast} & (c^{00}_{\mu \tau})^{*} & c^{00}_{\tau \tau} \\
  \end{array}
\right).
\end{equation}

The kind of corrections one gets in such a model are similar to the ones arising in a model that includes the possibility of Non Standard Neutrino Interactions (NSI), that can assume the form:
\begin{equation}
H_{NSI} = \sqrt{2} \, G_F \, N_e \times
\left(
  \begin{array}{ccc}
    \epsilon^{00}_{ee} & \epsilon^{00}_{e \mu} & \epsilon^{00}_{e\tau} \\
    (\epsilon^{00}_{e \mu})^{\ast} &  \epsilon^{00}_{\mu \mu} & \epsilon^{00}_{\mu \tau} \\
    (\epsilon^{00}_{e \tau})^{\ast} & (\epsilon^{00}_{\mu \tau})^{*} & \epsilon^{00}_{\tau \tau} \\
  \end{array}
\right)
\, ,
\end{equation}
with the formal correspondence $a^0_{\alpha \beta} = \sqrt{2} G_F N_e \, \epsilon_{\alpha \beta}$.
The values of the non diagonal matrix elements, that would contribute to lepton flavor violation, are of course limited by the data of oscillation experiments, like the SuperKamiokande atmospheric neutrino data~\cite{SK-atmo}, which put the following limits (in units of GeV): $|a^0_{e\mu}| \leq 2.5 \times10^{-23}; \, |a^0_{e\tau}| \leq 5 \times 10^{-23} ; |a^0_{\mu \tau}| \leq 8.3 \times 10^{-24}$.

In~\cite{LBL-LIV-indiani} the impact of the different CPT and LIV coefficients on the neutrino flavor oscillation probabilities and consequently on the $\nu_{\mu} \to \nu_{e}$ and $\nu_{\mu} \to \nu_{\mu}$
channels for the two LBL experiments T2K and No$\nu$A and the experiments' sensitivity to these coefficients are analyzed.
The main outcome of the analysis is that the $a^0_{e \mu}$ and $a^0_{e \tau}$, together with the matrix diagonal
element $a^0_{e e}$ are relevant for the $\nu_{\mu} \to \nu_e$ appearance channel, whereas the $\nu_{\mu}$
survival probability is influenced (but in a weaker way) by $a^0_{\mu \tau}$. Generally speaking the upper limits
obtained on these coefficients from the analysis of the T2K experiment
are almost one order of magnitude lower than the corresponding ones derived by the atmospheric neutrino data.
The limits become more stringent in the NO$\nu$A case and even more if one considers a ``combined T2K + NO$\nu$A''
analysis, but in any case they do not reach the level of SuperKamiokande limits.
The study performed in~\cite{LBL-LIV-indiani}, for the NO$\nu$A case also shows  that the CPT violating LIV
corrections can affect quite significantly the CP violation and mass ordering studies at LBL experiments, increasing
or decreasing their sensitivity according to the values of the LIV parameters phases.

A similar study~\cite{LIV-DUNE} has also been performed  for the case of the DUNE experiment~\cite{DUNE}.The~authors investigated the impact of potential $CPT$-violating LIV corrections on the experiment sensitivity for
the measurement of the $\theta_{23}$ octant and $CP$ phases. They found that the presence of a non null single LIV
coefficient, even with a relatively small value ($|a_{e \mu}|$ or $|a_{e \tau}| = 5 \times 10^{-24}$)
could reduce the DUNE discrimination power, and in addition due to the destructive interference between the phenomenological
parameters one wants to extract from the analysis (that is the mixing angle, the usual Dirac CP violation phase
$\delta$
and the new $CP$ phase of dynamical origin $\phi_{e \mu}/\phi_{e \tau}$). The~reduction of the $\theta_{23}$ octant
resolution would, instead be less significant in case one considers both the LIV parameters ($a_{e \mu}$ and
$a_{e \tau}$) together in the analysis.
%

As already discussed, there are models in literature in which the modification of the traditional Lorentz invariance symmetry is introduced without breaking the fundamental $CPT$ invariance. They also predict interesting phenomenological consequences that can be tested by using  various experiments, both with natural and artificial neutrino sources. We refer in particular to the HMSR model \cite{Torri, Antonelli-Miramonti-Torri, TorriPhD, Torri3, nostro-ifae}, in~which the LIV has a geometric kinematic origin and does not break space-time isotropy.  This feature is very important for  phenomenological analyses because it guarantees the possibility of looking for isotropic corrections to the ``traditional`` experimental signature, without the need of fixing any preferred reference frame.

The MDRs cause a variation in the neutrino propagation and, consequently,
a modification in the neutrino oscillation probabilities, which has been explicitly computed for  different neutrino flavors in~\cite{Antonelli-Miramonti-Torri,TorriPhD, Torri3}.
The LIV-induced corrections present an energy dependence different from the one of the
``usual'' leading term (they are proportional to $L \times E$, instead of  $\dfrac{L}{E}$). This introduces a partial deformation of the oscillating neutrino transition probability and of the corresponding
 experimental spectrum for various neutrino interaction channels.
 This effect could in principle be observed in experiments involving high values of the factor
$L \times E$, that is the product of the neutrino energy E and of the baseline L between
the neutrino production and interaction points. The~amplitudes of these corrections in the model considered would be proportional to $\delta f_{ij}$, which is the difference between the LIV coefficients
$f_i$ and $f_j$, correcting the dispersion relations for the two neutrino species $\nu_i$ and $\nu_j$
involved in the process under investigation.

 As shown explicitly in~\cite{Antonelli-Miramonti-Torri,TorriPhD, Torri3}, the LIV-induced variations in the oscillation
 probabilities could be relevant (of the order of some percent) for values of $\delta f_{ij}$ of the order
 of $10^{-23}$. Even for much lower values of  LIV coefficients
 $\delta f_{ij} \simeq 10^{-26}-10^{-27}$, compatible with a conservative interpretation of the limits derived (for different LIV models) by the analysis of SuperKamiokande atmospheric neutrino data~\cite{SK-atmo}, one can hope to observe the effects of the oscillation probability corrections, at least for experiments involving high or very high neutrino energies.
A first analysis of some of the most interesting experimental opportunities has been
performed~\cite{Antonelli-Miramonti-Torri,TorriPhD, Torri3, nostro-ifae, nostro-algeria}. Further analyses
can be done about the neutrino telescope (ANTARES~\cite{Antares}, KM3NeT~\cite{KM3NeT}, and
IceCube~\cite{Ice-Cube-LIV-potentiality, IceCube-high, IceCube-high-2}) detection of high-energy neutrinos (with E in the range between TeV and PeV),
the Auger study of cosmic neutrinos with $E > EeV$~\cite{Auger-neutrinos}, and the potential analysis
of high-energy atmospheric neutrinos at the JUNO experiment~\cite{JUNO, Antonelli-Miramonti-Ranucci}.
Some of these studies are in progress and will be published soon.

Finally, to complete the theoretical and phenomenological framework, it is worthwhile to recall that an interesting general analysis of the limits that can be derived for LIV from the study of high energy cosmic neutrinos can also be found in~\cite{Carmona}.

\vspace{6pt}

\textbf{Author contributions}
Writing—original draft: M.D.C.T., V.A. and L.M.; review and editing:
M.D.C.T. All authors contributed equally to this work. All authors have read and agreed to the published version
of the manuscript.

\textbf{Funding}
This work was supported by the Fondazione Fratelli Giuseppe Vitaliano, Tullio e Mario Confalonieri - Milano, financing a post doctoral fellowship.

\textbf{Conflicts of interest}
The authors declare no conflict of interest.

\end{document}